\documentclass{article}

\usepackage{arxiv}

\usepackage[utf8]{inputenc} 
\usepackage[T1]{fontenc}    
\usepackage{hyperref}       
\usepackage{url}            
\usepackage{booktabs}       
\usepackage{amsfonts}       
\usepackage{nicefrac}       
\usepackage{microtype}      
\usepackage{lipsum}
\usepackage{graphicx}
\usepackage{natbib}
\usepackage{amsmath}
\usepackage{newtxtext}
\usepackage{newtxmath}
\numberwithin{equation}{section}

\usepackage{algorithm,algcompatible}
\usepackage{algorithmicx}%
\usepackage{algpseudocode}%
\usepackage{caption}

\usepackage{bbm}
\usepackage{color}

\usepackage{hyperref}
\hypersetup{
    colorlinks = true,
    urlcolor   = black,
    citecolor  = blue,
}

\title{Forcing-informed resolvent analysis: Identification of input-output relations in self-sustained flows}

\usepackage{authblk}

\newbox{\orcid}\sbox{\orcid}{\includegraphics[scale=0.06]{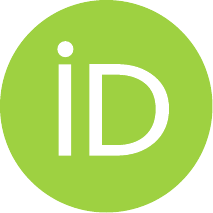}} 
\author[1]{%
	\href{https://orcid.org/0000-0002-0676-0487}{\usebox{\orcid}\hspace{1mm}
    Yuta Iwatani\thanks{\texttt{Corresponding author:yuta.iwatani@tohoku.ac.jp}}}%
}
\author[2]{%
	\href{https://orcid.org/0000-0002-3762-8075}{\usebox{\orcid}\hspace{1mm}
    Kunihiko Taira}%
}
\author[1]{%
	\href{https://orcid.org/0000-0002-0515-4071}{\usebox{\orcid}\hspace{1mm}
    Soshi Kawai}
}
\affil[1]{Department of Aerospace Engineering\\ Graduate School of Engineering\\ Tohoku University, Sendai, Miyagi\\ 980-8579, Japan}
\affil[2] {Department of Mechanical and Aerospace Engineering, University of California, Los Angeles, CA 90095, USA}

\begin{document}

\maketitle

\begin{abstract}
We present a forcing-informed (FI) resolvent analysis framework to identify input-output relations for statistically stationary self-sustained unsteady flows.
The central idea of this method is to inform the resolvent operator about the spatiotemporal structures of the nonlinear terms that act as exogenous forcing with respect to the mean flow.
To construct the FI resolvent operator, we estimate the basis vectors for the input subspace spanned by forcing snapshots and, similarly, for the output subspace, from simulation data.
The extracted FI response and forcing modes are expressed through the estimated bases of the output and input subspaces, respectively, and the singular values of the FI resolvent operator correspond to the actual output amplitudes. 
These properties ensure that the extracted modes are consistent with the actual self-sustained flow fields.
Additionally, the forcing snapshots can be used to construct the linear operator, enabling a fully data-driven FI resolvent analysis.
The proposed framework is validated using the Stuart-Landau oscillator and demonstrated for a two-dimensional cylinder wake and a three-dimensional transitional boundary layer.
We successfully identify the gains and the corresponding pairs of forcing and response modes, even at frequencies where the nonlinear amplification mechanism is crucial.
Furthermore, leveraging the balance between the time-averaged energy amplification/attenuation by the linear operator and nonlinear forcing, we introduce a nonlinear energy transfer map that identifies the spatial domains where the extracted forcing mode injects or removes fluctuation energy, thereby providing key physical insight into the self-sustaining mechanisms.
\end{abstract}

\section{Introduction}
Self-sustaining mechanisms underlying a wide range of unsteady flows have been one of the most fundamental questions in fluid mechanics due to their scientific and engineering importance.
To date, numerous works have been performed to study the self-sustaining mechanisms, in wall turbulence (e.g., \citealt{Jimenez1991-oa, Waleffe1997-rn,Hwang_2015,Motoori2021-rw}), cavity flows (e.g., \citealt{Rockwell1978-oa,singh2025effects}), and transonic airfoil buffet flows (e.g., \citealt{Lee2001-fa, Kojima2020-bc, Iwatani2023-az}).
However, identifying such self-sustaining mechanisms is challenging due to the complex, multiscale nature of fluid dynamics, even with the knowledge of the governing equations and access to a high-fidelity dataset from large-eddy simulation (LES) or direct numerical simulation (DNS).
To this end, this study proposes an input-output analysis framework that utilizes information on nonlinear forcing in the feedback mechanism for resolvent analysis and ultimately develops a fully data-driven formulation for statistically stationary flows, making full use of high-fidelity LES or DNS datasets.

The motion of fluids is governed by an infinite-dimensional Navier-Stokes (NS) equation and is numerically solved on numerous grid points that sometimes exceed a billion, even for an industrial application as presented by \citet{Asada2023-jz}. 
In such cases, the flow description is too high-dimensional to identify the physically important locations and time scales, and the associated data analysis requires high computational cost.
On the other hand, the essential dimensionality of the dynamics can be significantly reduced due to the dissipative nature of the NS equation, which is particularly useful for dimensionality reduction.
In fact, many data-driven approaches have successfully constructed reduced-order models for a variety of seemingly high-dimensional and complicated flows, too numerous to list here, including modal analyses \citep{Taira2017-kn,Taira2020-rh,Schmidt2020-qa}, autoencoder machine learning models \citep{Linot2023-aa,Page2024-ex,Fukami2025-vs}, spectral submanifold \citep{Kaszas2024-hb},
phase-amplitude reduction \citep{fukami2024data} and parametric reduced order model \citep{Nakamura2024-fz,Sato_Schmidt_2025}.
These studies motivate the use of data-driven approaches for extracting input-output relations in nonlinear self-sustaining flows.

Resolvent analysis \citep{Trefethen1993-pv,Jovanovic2005-rq,McKEON2010-xc,Rolandi2024-jr} is a powerful framework for uncovering input-output relations based on the linearized NS equations.
The resolvent analysis can identify the pair of the input and corresponding output modes, called the forcing and response modes, respectively, via the singular value decomposition (SVD) of the resolvent of the linearised NS operator about the base flow of interest.
The resolvent analysis can also be generalized by considering the input and output windows in analogy to the state-space model used in control theory \citep{Jovanovic2005-rq, Jeun2016-gi,Jovanovic2021-cu}.

In the literature on resolvent analysis, the formulation based on the feedback mechanism \citep{Farrell1993-ui,McKEON2010-xc} has extended the utility of the resolvent analysis to turbulent base flows.
In the feedback mechanism, the nonlinear terms with respect to the perturbation components are viewed as a self-induced feedback controller that sustains the flow oscillations, and this feedback forcing is regarded as the (harmonic) exogenous forcing for the linear system.
With the formulation, the resolvent analysis has been used for flow control \citep{Luhar2014-xo, Yeh2019-rn,Ribeiro_Taira_2024}, flow modeling \citep{Gomez2016-el, Lesshafft2019-kh}, and elucidating the physics of the self-sustaining mechanism \citep{Nogueira2021-kz,bae2021nonlinear}.

However, it is reported that the modes revealed by resolvent analysis are sometimes not observed in the actual nonlinear self-sustained flow field in numerical simulations or experiments due to the lack of full information on nonlinear forcing \citep{Taira2017-kn}; the linear operator holds the information of the mean flow distorted by the Reynolds stresses, but it does not hold the information on fluctuation components of the nonlinear forcing, often referred to as the colour of forcing \citep{Zare2017-tc}.
The issue of inconsistency between the extracted modes and the actual flow is not limited to nonlinear self-sustained flows.
\citet{Kamal2022-vz} pointed out that the worst-case disturbance obtained in the linear receptivity analysis cannot be physically realizable, and proposed a framework that restricts the input forcing to a set of physically realizable disturbances.
For statistically stationary flows, incorporating nonlinear forcing statistics enables us to obtain the response mode that appears in nonlinear flows \citep{Zare2017-tc, Towne2018-zc} and also enhances the performance of resolvent-analysis-based flow control \citep{Jung2025-lq}.

Recognizing the importance of nonlinear forcing information, recent efforts analyze the forcing statistics themselves \citep{Chen2025-im} and attempt to incorporate the nonlinear forcing terms into the resolvent analysis formulation.
Broadly speaking, these studies can be classified into four approaches.
The first approach, which we will also adopt, is to directly compute all nonlinear forcing terms \citep{Towne2020-xz,Nogueira2021-kz}. 
\citet{Nogueira2021-kz} showed that the cross-spectral density (CSD) of the output is recovered from the resolvent operator and the CSD of the nonlinear forcing. 
They subsequently examined the essential forcing components in the minimum channel flow to understand its self-sustaining process.
The second approach is to incorporate the nonlinear forcing effect into the linear operator via the eddy viscosity modeling \citep{Hwang2010-ea, Morra2019-tg,Morra2021-jk,Pickering2021-pt,Symon2021-gf, Holford2023-gq, Fan2024-cg,Ying2024-da} as in the linear stability analysis of turbulent flows \citep{Del_ALAMO2006-xx}.
The third approach is to attempt to close the feedback-loop formulation using only the output modes by explicitly modeling the triadic interactions associated with the quadratic nonlinear forcing terms \citep{Rosenberg2019-yr, Barthel2021-an}.
The last fourth approach is to incorporate the nonlinear forcing feedback as structured uncertainty \citep{Liu2021-ex,song2026structured}.

In addition to the feedback-mechanism-based formulation for the treatment of the nonlinear forcing, there have been several attempts to directly incorporate the nonlinearity of the NS equation into input-output analysis, including harmonic resolvent analysis \citep{Padovan2020-jx,Padovan2022-kt} based on the harmonic transfer function defined for a linear time-periodic system \citep{Wereley1991-sz} and a nonlinear input-output analysis method based on harmonic-balanced NS equations \citep{Rigas2021-ac}.
There is also the mean resolvent analysis \citep{Leclercq2023-sx} based on the mean resolvent operator that describes the time-averaged linear response to any given forcing and can be viewed as an implicit extension of the resolvent operator manipulated, for instance, by eddy-viscosity modeling of nonlinear forcing (the second approach).
As extensions of the resolvent analysis for transient growth, there are the nonlinear optimum perturbation analysis \citep{Pringle2010-ju,Huang2020-nw} and its extension to external forcing \citep{Taniguchi2026-if}.
Although numerous input-output analysis methods that incorporate the nonlinear forcing have been developed, there is no method that simply extracts input-output relations that are theoretically grounded to exist in the actual self-sustained flow fields.

\begin{figure}
    \centering
    \includegraphics[width=0.9\linewidth]{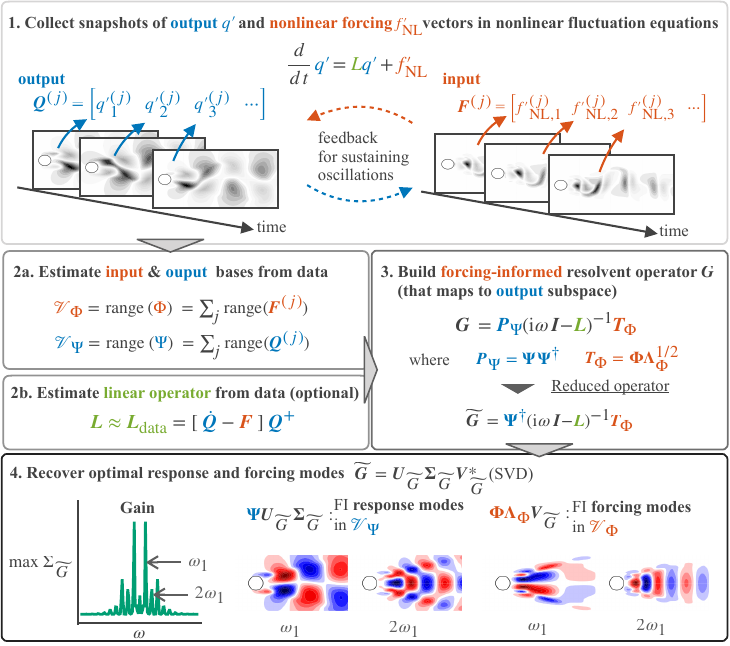}
    \caption{Schematic of workflow of the present forcing-informed (FI) resolvent analysis, shown for an example of a two-dimensional cylinder case. 
    Detailed mathematical formulations are presented in \S\ref{sec:forcinginformedresolvent} and \S\ref{sec:numalg}.
    }
    \label{fig:summary}
\end{figure}

Therefore, to identify the input-output relations in self-sustained unsteady flows, this study presents a forcing-informed (FI) resolvent analysis framework, based on the concept of feedback-mechanism formulation \citep{Farrell1993-ui,McKEON2010-xc}, as summarized in Fig.~\ref{fig:summary}.
The central idea in this method is to inform the conventional resolvent operator of the spatio-temporal structure of the nonlinear forcing terms crucial for sustaining mechanisms.
To do so, we collect snapshots of nonlinear forcing terms (step 1) and learn the basis vectors of the input subspace spanned by them (step 2a). 
These basis vectors are used to derive the FI resolvent operator (step 3), a core quantity of this method.
The FI resolvent operator enables the extraction of the forcing and response modes, which always exist in the self-sustained flows, even at frequencies where the nonlinear amplification mechanism dominates and is not captured by conventional resolvent analysis. 
As shown in step 4, the modes at the second harmonic of vortex shedding behind a cylinder are extracted.
Furthermore, the nonlinear forcing snapshots are used to estimate the linear operator in a data-driven manner (step 2b), thereby enabling a fully data-driven formulation of this method. 
We show that the present FI forcing resolvent analysis can also be viewed as an extension of the data-driven resolvent analysis \citep{Herrmann2021-fk} for the statistically stationary nonlinear flows.

The rest of this paper is organized as follows.
We describe the theoretical framework of the FI resolvent analysis in \S\ref{sec:2} and present the details of the numerical algorithms in \S\ref{sec:numalg}.
In \S\ref{sec:summary}, the proposed method is summarized briefly while revisiting Fig.~\ref{fig:summary}. 
We demonstrate the capabilities of the present FI modes in \S\ref{sec:results}. 
We begin with a proof-of-concept using the Stuart-Landau oscillator, then move to a simple two-dimensional cylinder wake exhibiting self-sustained vortex shedding. 
Finally, we use the present approach to study the three-dimensional (3D) late stage of a transitional boundary layer, as a practical case.

\section{Forcing-informed (FI) resolvent analysis}\label{sec:2}
\subsection{Nonlinear fluctuation equations for statistically stationary self-sustained flows}\label{sec:2.1}
Let us consider the spatially discretised compressible Navier--Stokes equations for the state vector $q = [\rho_1, \boldsymbol{m}^t_1, E_1,\cdots, \boldsymbol{m}^t_{N_g}, E_{N_g}]^T \in \mathbb{R}^{N_q}$ in a general form,
\begin{align}
\frac{d q}{d t} = N(q) ,
\label{eq:NS}
\end{align}
where $t$ is time, $\rho$ is density, $\boldsymbol{m}=[m_x,m_y,m_z]^T$ is momentum vector, $E$ is total energy, superscript $T$ denotes the transpose of a real vector, and $N_q$ is the product of the number of total grid points $N_g$ and the number of variables.
Alongside, considering the Reynolds decomposition of the state vector, i.e., $q(t)=\bar{q}+q'(t)$ with the time averaging operator being defined as
\begin{align}
    \overline{[\cdot]} = \lim_{t'\rightarrow\infty} \frac{1}{t'} \int_{0}^{t'} [\cdot](t)  dt,
\end{align}
we obtain the Reynolds-averaged governing equations:
\begin{align}
 0 =\overline{N(q)} =  N(\overline{q})+ \overline{F({q'}^{n})}  ~(n\ge 2).
 \label{eq:RANS}
\end{align}
For the second equality in \eqref{eq:RANS}, we have used the relation
\begin{align}
    N(\overline{q}+q')=N(\bar{q}) + L_{\bar{q}}(q')+F_{\mathrm{NL}}({q'}^{n}).
    \label{eq:ReynoldsDecomposition}
\end{align}
Here, $N(\bar{q})$, $L_{\bar{q}}(q')$, $F_{\mathrm{NL}}({q'}^{n})$ collect the constant terms,  linear terms, and nonlinear terms with respect to the fluctuation of the state vector $q'$, respectively.
Only the linear terms $L_{\bar{q}}(q')$ can be expressed in the form of a matrix-vector product as $L_{\bar{q}}(q')=\boldsymbol{L}q'~(\boldsymbol{L}\in \mathbb{R}^{N_q\times N_q})$ and become zero by taking the time average.
Subtracting the Reynolds-averaged equations \eqref{eq:RANS} from the governing equations \eqref{eq:NS} yields the nonlinear fluctuation equation, written as
\begin{align}
    \frac{dq'}{d t} = \boldsymbol{L} q' + f'_{\rm NL},
    \label{eq:NLpert}
\end{align}
where
\begin{equation}
    f'_{\rm NL} = F_{\mathrm{NL}}({q'}^n)-\overline{F_{\mathrm{NL}}({q'}^n)} .
    \label{eq:NLforcing}
\end{equation}
The components of vector $f'_{\mathrm{NL}}$ for compressible NSs are derived in appendix \ref{app:forcingterm}.
By definition, all the terms in \eqref{eq:NLpert} are zero mean (centered).
The fluctuation equations \eqref{eq:NLpert} are applicable to the statistically stationary system or the linearised system around the equilibrium point ($f'_{\rm{NL}}$ is negligibly small), but not applicable to cases where an appropriate mean flow $\overline{q}$ cannot be defined.

Importantly, the nonlinear fluctuation equation~\eqref{eq:NLpert} is viewed as the linear system $\boldsymbol{L}$ with the exogenous forcing $f'_{\rm{NL}}$ as a model of a feedback controller to sustain the oscillation of $q'$ \citep{Farrell1993-ui,McKEON2010-xc}.
The forcing $f'_{\rm{NL}}$ should attenuate the fluctuation energy for an unstable linear system $\boldsymbol{L}$, while it should act as a driving force for a stable linear system $\boldsymbol{L}$.
In either case, the nonlinear forcing terms are crucial for sustaining the oscillation $q'$.
To consider the role of forcing term $f'_{\rm{NL}}$ for statistically stationary self-sustained flows, we consider the (time-averaged) evolution of disturbance energy $E_{q'}$, defined by a weighted inner product $E_{q'}(t)\coloneq (1/2)\langle q', q'\rangle_W:=(1/2){q'}^*\boldsymbol{W}q'$ with $\boldsymbol{W}$ being positive semi-definite matrix.
The evolution of the fluctuation energy $E_{q'}$ is  governed by
\begin{align}
    {\frac{dE_{q'}}{dt}} &=
    \frac{1}{2}\left\langle \frac{dq'}{dt},q'\right\rangle_W
    +\frac{1}{2}\left\langle q',\frac{dq'}{dt}\right\rangle_W\nonumber\\
    &= {\left\langle q', \frac{dq'}{dt}\right\rangle_W
    } 
    \nonumber \\
    &={\left\langle q',\boldsymbol{L}q'\right\rangle_W} +{\langle q',f'_{\rm NL}\rangle_W} ~(\because \eqref{eq:NLpert}).
\end{align}
Given that $\overline{dE_{q'}/dt}=0$ for statistically stationary flows, the time-averaged equation is obtained as
\begin{align}
    \overline{\frac{dE_{q'}}{dt}} 
    =\overline{\langle q',\boldsymbol{L}q'\rangle_W} + \overline{\langle q',f'_{\rm NL}\rangle_W} 
    =0.
   \label{eq:timeave_evolution_of_Eq'}
\end{align}
This equation \eqref{eq:timeave_evolution_of_Eq'} describes the time-averaged energy balance between the linear amplification/attenuation mechanism $\overline{\langle q',\boldsymbol{L}q'\rangle_W}$ and the nonlinear mechanism $\overline{\langle q',f'_{\rm{NL}}\rangle_W}$.
In this balance relation, the term $\overline{\langle q',f'_{\rm{NL}}\rangle_W}$ is associated with the inter-scale nonlinear energy transfer \citep{Jin2021-ge}.
This time-averaged energy balance is one of the key concepts in the proposed framework, which aids the physical interpretation of the extracted input-output modes, as discussed later.

We are interested in how such a nonlinear forcing term $f'_{\rm{NL}}$ behaves in space and time to understand its role in the self-sustaining mechanisms.
Therefore, this study aims to extract the important component in the nonlinear forcing term $f'_{\rm{NL}}$ and the corresponding response $q'$, with the aid of their time-series data from high-fidelity numerical simulations.

\subsection{Conventional resolvent analysis}
Before discussing the treatment of the nonlinear forcing terms, let us consider the conventional, particularly, Fourier-transform-based, resolvent analysis to clarify the necessity of information about the spatiotemporal correlation of the nonlinear forcing term $f'_{\mathrm{NL}}$.
For the statistically stationary system, the Fourier transform of \eqref{eq:NLpert} yields the relation at a frequency $\omega$:
\begin{align}
    \hat{q}(\omega) = \boldsymbol{R}(\omega) \hat{f}({\omega}),
    \label{eq:IO}
\end{align}
where $\boldsymbol{R}=[ - \mathrm{i}\omega \boldsymbol{I}-\boldsymbol{L}]^{-1}$ is the so-called resolvent operator and $\boldsymbol{I}$ is the identity matrix. 
In the resolvent analysis, we consider the following maximization problem of the energy gain between the forcing and response as
\begin{align}
    G^2(\omega) = \max_{\hat{f}(\omega)\ne0}\frac{\|\hat{q}(\omega)\|_W^2}{\|\hat{f}(\omega)\|_W^2} = \max_{\hat{f}(\omega)\ne0}\frac{\|\boldsymbol{R}(\omega)\hat{f}(\omega)\|_W^2}{\|\hat{f}(\omega)\|_W^2},
    \label{eq:Gain}
\end{align}
where $\|\cdot\|_W = \sqrt{\langle \cdot,\cdot\rangle_W}$ is the energy norm.
Although the use of different weighted norms for input and output (the denominator and numerator of \eqref{eq:Gain}) is possible, we consistently use the same weighted norm $\boldsymbol{W}$ to define the inner product $\langle q',f'_{\rm{NL}}\rangle_W$ in \eqref{eq:timeave_evolution_of_Eq'}.
The maximization problem \eqref{eq:Gain} is solved using the SVD of the resolvent operator $\boldsymbol{R}$, expressed as
\begin{align}
    \boldsymbol{R}(\omega) = \sum_{j=1}^{N_R} u_j(\omega) \sigma_j(\omega) \langle v_j(\omega), ~\cdot~\rangle_W,
    \label{eq:SVDofR}
\end{align}
where $N_R$ is the rank of matrix $\boldsymbol{R}$, $\sigma_j\in\mathbb{R}$ is the $j-$th greatest singular value, i.e., $\sigma_1\ge\sigma_{2}\ge\cdots\ge\sigma_{N_R}>0$, and $u_j, v_j\in \mathbb{C}^{N_q}$ are the corresponding left and right singular vectors, respectively.
Here, the singular vectors satisfy the following orthogonal conditions:
\begin{align}
    \langle u_j(\omega),u_k(\omega)\rangle_W =    \langle v_j(\omega),v_k(\omega)\rangle_W =\delta_{jk},
\end{align}
where $\delta_{jk}$ is the Kronercker delta.
In numerical computation, we may obtain such singular vectors as $u_k=\boldsymbol{Z}^{-1}_w u_k^w$ and $v_k=\boldsymbol{Z}^{-1}_w v_k^w$.
Here, $\boldsymbol{Z}_W$ is the outcome of the Cholesky decomposition of the weight matrix $\boldsymbol{W}=\boldsymbol{Z}_w^*\boldsymbol{Z}_w$ and $u_k^w$ and $v_k^w$ are obtained by the SVD of $\boldsymbol{Z}_W\boldsymbol{R} \boldsymbol{Z}_W^{-1}=\sum_{j}u_j^w\sigma_j \langle v_j^w,\cdot\rangle_2$, where $\langle\cdot,\cdot\rangle_2$ denotes the Euclidean inner product.
Once we obtain the SVD results, selecting $\hat{f}=v_1$ in \eqref{eq:Gain} leads to the maximum gain $G=\sigma_1$ and the corresponding response $\hat{q}=u_1\sigma_1$.
Not only the most amplified input-output pair, but the $j$-th amplified pair can also be extracted by selecting $\hat{f}=v_j$.
In this sense, the left singular vector $u_j$ and the right singular vector $v_j$ are called the response mode and the forcing mode, respectively, and the singular value $\sigma_j$ is called the gain.

However, the resolvent operator $\boldsymbol{R}$ relies only upon the information of the linear operator $\boldsymbol{L}$, and hence does not possess the full information (spatiotemporal structure) of the fluctuating nonlinear forcing term $f'_{\rm{NL}}$ that plays a crucial role in the self-sustaining mechanism.
The temporal structure of the nonlinear forcing term is reflected as the magnitude of the Fourier amplitude $|\hat{f}|$ while the spatial structure is reflected as the vector shape $\hat{f}$.
In what follows, we point out the issues with neglecting the forcing information from three perspectives, including discussions in previous studies.

First, the cross-spectrum density (CSD) tensor of the output signal $\boldsymbol{S}_{\mathrm{qq}}:=E[\hat{q}\hat{q}^*]$, where $E[\cdot]$ is the expectation (ensemble averaging) operator, cannot be recovered \citep{Zare2017-tc,Towne2018-zc,Morra2019-tg,Towne2020-xz,Nogueira2021-kz}. 
From \eqref{eq:IO}, we may derive 
\begin{align}
    \boldsymbol{S}_{\mathrm{qq}}(\omega)= \boldsymbol{R}(\omega) \boldsymbol{S}_{\mathrm{ff}}(\omega) \boldsymbol{R}^*(\omega),
    \label{eq:CSD}
\end{align}
where $\boldsymbol{S}_{\mathrm{ff}}$ is the CSD tensor of the forcing term $\hat{f}_\omega$.
This indicates that the $\boldsymbol{S}_{\mathrm{qq}}$ cannot be estimated solely from the resolvent operator $\boldsymbol{R}$ unless $\boldsymbol{S}_{\mathrm{ff}}=\boldsymbol{I}$, i.e., the forcing signal is the spatially and temporally uncorrelated (white noise) and has the unit-norm, but this is not true in general self-sustained nonlinear flows. 
This means that the resolvent response mode and the spectral proper orthogonal (SPOD) mode \citep{Lumley1967-ce,Schmidt2020-qa} are identical only for the system perturbed by unit-norm white noise \citep{Towne2018-zc} since the eigendecomposition of $\boldsymbol{S}_{\rm{qq}}$ results in the SPOD mode of $\hat{q}_\omega$.

Second, although the conventional resolvent analysis ranks the most dominant forcing and response pair based on the gain (singular value), this ordering is no longer theoretically supported for the spatiotemporally correlated forcing.
Combining \eqref{eq:IO} and \eqref{eq:SVDofR} leads to
\begin{align}
    \hat{q}(\omega) = \sum_{j=1}^{N_R} u_j(\omega) \sigma_j(\omega) \langle v_j(\omega),\hat{f}(\omega)\rangle_W, 
\end{align}
which indicates that the product of the singular value $\sigma_j$ and the inner product $\langle v_j,\hat{f}\rangle_W$ determines the expansion coefficient (amplitude) of $u_j$. 
This in turns means that the situation $|\sigma_j\langle v_j,\hat{f}\rangle_W|\le|\sigma_{j+1}\langle v_{j+1},\hat{f}\rangle_W|$ is possible especially at the frequency in which the linear amplification mechanism is inactive while the nonlinear one is active (i.e.,  $\sigma_1$ is small while $|\hat{f}|$ is large).
This property makes it less justifiable to examine the response and forcing modes associated with the leading gain under correlated forcing conditions.

Third, in terms of linear algebra, the maximization problem \eqref{eq:Gain} searches the entire vector space $\mathbb{C}^N$ for the optimum forcing. 
This indicates that there is no theoretical guarantee that the optimum forcing (the right singular vector $v_1$) reflects the spatial coherence of the nonlinear forcing term.
In addition, the search space is restricted onto the unit sphere due to the invariance of the Rayleigh quotient \eqref{eq:Gain} with respect to the scaling of input $\hat{y}=c\tilde{y}$.
This means that the optimal forcing $v_1$ does not reflect the temporal coherence captured by the amplitude $|\hat{f}|$. 
As such, the optimal forcing selected via the SVD of $\boldsymbol{R}$ does not always lie in the vector subspace spanned by the nonlinear forcing term $\hat{f}$, and such possibly physically inconsistent optimal forcing can lead to the discrepancy between the response modes and actual flow field obtained from numerical simulations or experiments.

Therefore, these concerns call for treatments of the spatiotemporally correlated forcing. 
In particular, to identify the input-output relations from self-sustained unsteady flows, the third statement necessitates choosing the optimal forcing from the vector subspace spanned by the actual correlated nonlinear forcing $\hat{f}$, rather than from the entire space $\mathbb{C}^N$.
This requirement provides the key insight behind the proposed framework, which is detailed in the next section.

\subsection{Forcing-informed resolvent operator}\label{sec:forcinginformedresolvent}
Based on previous discussion, we propose a method to incorporate the information of the forcing term $f'_{\rm{NL}}$ into the resolvent operator.
As in the previous section, the core idea of the proposed framework is to introduce input and, similarly, output subspaces, each spanned by snapshot data from numerical simulations, and to estimate their bases.
The estimated basis of the input space is used to inform the forcing $f'_{\rm{NL}}$ of the conventional resolvent operator, so that the `forcing-informed' resolvent operator, which we will derive as \eqref{eq:FIresolventoperator} in this section, captures the input-output relations in nonlinear self-sustained flows. Similarly, the basis of the output space is used to ensure that the response mode lies within the actual flow field obtained from numerical simulation.

First, we define the input and output subspaces.
To numerically obtain bases for the input and output spaces, we consider the time-series datasets of output $q'$ and input $f'$ obtained from multiple numerical simulation runs (or realizations). 
For statistically stationary flow, obtaining multiple realizations is equivalent to sampling time-series data from different time segments, as in Welch's method \citep{welch2003use}.
The input subspace $\mathcal{V}_\Phi$ and output subspace $\mathcal{V}_\Psi$ spanned by their time-series datasets are introduced as 
\refstepcounter{equation}
\begin{equation*}
\mathcal{V}_{\Phi} =
\operatorname{span}
\left\{
{f}_{\mathrm{NL},m}^{\prime (n)}
\right\}_{m=1,n=1}^{M,N_p},
\tag{$\theequation{\mathit{a}}$}
\label{eq:input_space}
\end{equation*}
\begin{equation*}
\mathcal{V}_{\Psi} =
\operatorname{span}
\left\{
{q'}_{m}^{(n)}
\right\}_{m=1,n=1}^{M,N_p},
\tag{$\theequation{\mathit{b}}$}
\label{eq:output_space}
\end{equation*}
where $M$ is the number of snapshots, $N_p$ is the number of realizations, and $q^{\prime(n)}_{m}$ and $f^{\prime(n)}_{\mathrm{NL},m}$  indicate the $m$-th output and nonlinear forcing snapshots of the $n$-th realization, respectively.
Performing the Fourier transform for each realization and collecting the outcome at each frequency $\omega$, we construct the matrices written as
\refstepcounter{equation}
\begin{equation*}
\widehat{\boldsymbol{F}}(\omega)=
\begin{bmatrix}
    \hat{f}^{(1)}(\omega)&\dots&\hat{f}^{(N_p)}(\omega)
\end{bmatrix}\in\mathbb{C}^{N_q\times N_p}, , 
\label{eq:forcfreqmat}
\tag{$\theequation{\mathit{a}}$}
\end{equation*}
\begin{equation*}
\widehat{\boldsymbol{Q}}(\omega)=
\begin{bmatrix}
    \hat{q}^{(1)}(\omega)&\dots&\hat{q}^{(N_p)}(\omega)
\end{bmatrix}\in\mathbb{C}^{N_q\times N_p}, 
\tag{$\theequation{\mathit{b}}$}
\end{equation*}
where $\hat{f}^{(j)}$ and $\hat{q}^{(j)}$ are the Fourier coefficients of input and output obtained from the snapshots of $j$-th realisation.
The orientations of column vectors of $\widehat{\boldsymbol{F}}$ and $\widehat{\boldsymbol{Q}}$  capture the spatial structures of the input $f'_{\rm{NL}}$ and the output $q'$.
Such column spaces of the matrices span the input and output subspaces at frequency of $\omega$,  respectively written as 
\refstepcounter{equation}
\begin{equation*}
\mathcal{V}_{\Phi_\omega}(\omega)=\operatorname{Col}(\widehat{\boldsymbol{F}}(\omega))=\operatorname{span}\{\hat{f}_j(\omega)\}_{j=1}^{N_p},
\tag{$\theequation{\mathit{a}}$}
\end{equation*}
\begin{equation*}
\mathcal{V}_{\Psi_\omega}(\omega)=\operatorname{Col}(\widehat{\boldsymbol{Q}}(\omega))=\operatorname{span}\{\hat{q}_j(\omega)\}_{j=1}^{N_p},
\tag{$\theequation{\mathit{b}}$}
\end{equation*}
where $\operatorname{Col}(\cdot)$ denotes the column space of a matrix.
The linear combinations of the basis vectors of $\mathcal{V}_{\Phi_\omega}$ and $\mathcal{V}_{\Psi_\omega}$ for all the scales $\omega$ recover $\mathcal{V}_{\Phi}$ and $\mathcal{V}_{\Psi}$.

In general, the choice of bases of input and output subspaces is not unique, considering the rank factorization of a matrix as detailed in Appendix \ref{app:rankdecom}.
For brevity, we here consider the orthonormal basis vectors of the column spaces of $\widehat{\boldsymbol{F}}$ and $\widehat{\boldsymbol{Q}}$ as the basis vectors of the input and output subspaces, respectively.
We introduce the basis vectors of input and output subspaces at frequency $\omega$, respectively written as 
\refstepcounter{equation}
\begin{equation*}
 \boldsymbol{\Phi}(\omega)=
 \begin{bmatrix}
 \phi_1(\omega)&\phi_2(\omega)&\cdots&\phi_{N_\phi}(\omega)
 \end{bmatrix} \in \mathbb{C}^{N_q\times N_{\phi}},
 \tag{$\theequation{\mathit{a}}$}
\end{equation*}
 \begin{equation*}
 \boldsymbol{\Psi}(\omega)=
 \begin{bmatrix}
     \psi_1(\omega)&\psi_2(\omega)&\cdots&\psi_{N_p}(\omega)
 \end{bmatrix}\in \mathbb{C}^{N_q\times N_{\psi}}.
 \tag{$\theequation{\mathit{b}}$}
 \label{eq:basisvectors}
\end{equation*}
Here, $N_\Phi$ and $N_\Psi$ are the ranks of the matrices $\widehat{\boldsymbol{F}}$ and $\widehat{\boldsymbol{Q}}$, which are smaller than $ N_p$ (the number of the realizations) and $N_q$ (the length of the state vector). 
They are equivalent to the number of the basis vectors of the output and input subspaces, respectively.
These basis vectors satisfy the weighted orthonormal conditions 
\begin{align}
    \langle {\phi}_{i}(\omega), {\phi}_{j}(\omega)\rangle_W = \langle \psi_{i}(\omega), {\psi}_{j}(\omega)\rangle_W = \delta_{ij}.
    \label{eq:orthonormal}
\end{align}
Using these orthonormal basis vectors, the Fourier components in the learned input and output subspaces are expressed as follows:
\begin{subequations}
\begin{align}
    \hat{f}_{\Phi}(\omega) &= \underset{=:\boldsymbol{T}_{\Phi}}{\underbrace{\mathbf{\Phi}(\omega) \mathbf{\Lambda}_\Phi^{1/2}(\omega) }}\hat{a}(\omega), 
    \label{eq:basis_in_uspace_f}  
    \\[1ex]
    \hat{q}_{\Psi}(\omega) &= \mathbf{\Psi}(\omega) \mathbf{\Lambda}_\Psi^{1/2}(\omega) \hat{b}(\omega).
    \label{eq:basis_in_uspace_q} 
\end{align}
 \label{eq:basis_in_uspace}
\end{subequations}
Here, $\mathbf{\Lambda}_\Phi^{1/2}$ and $\mathbf{\Lambda}_\Psi^{1/2}$ are the diagonal matrices, each element expresses the amplitude of the input and output basis vector (obtained by the inner product between the dual basis and the actual input and output), and
$\hat{a},\hat{b}$ are normalized coefficient vectors, i.e., $\|\hat{a}\|_2=\|\hat{b}\|_2=1$, which determine the alignment among basis vectors.
Now, the information about the spatiotemporal structure of the forcing term $f'_{\rm{NL}}$ is captured in $\boldsymbol{T}_{\Phi}$ in (\ref{eq:basis_in_uspace_f}).

Using the learned basis vectors of input and output subspaces, we derive the FI resolvent operator.
By projecting \eqref{eq:IO} onto the output basis using the orthogonal projection matrix $\boldsymbol{P}_{\Psi}=\mathbf{\Psi}\mathbf{\Psi}^\dagger$, where $\mathbf{\Psi}^\dagger=\mathbf{\Psi}^*\boldsymbol{W}$ such that $\mathbf{\Psi}^\dagger\mathbf{\Psi}=\boldsymbol{I}$, and subsequently substituting (\ref{eq:basis_in_uspace_f}), we obtain the following relation at a frequency $\omega$:
\begin{align}
    \hat{q}_{\Psi}(\omega) = 
    {{\boldsymbol{P}_{\Psi}(\omega)\boldsymbol{R}(\omega)\boldsymbol{T}_\Phi}}(\omega) \hat{b}(\omega).
    \label{eq:IO_full}
\end{align}
Here, we refer to this newly defined matrix 
\begin{align}
    \boldsymbol{G}(\omega) :=  \boldsymbol{P}_{\Psi}(\omega)\boldsymbol{R}(\omega) \boldsymbol{T}_\Phi(\omega) \in \mathbb{C}^{N_q\times N_\psi},
    \label{eq:FIresolventoperator}
\end{align}
as the forcing-informed (FI) resolvent operator.
The FI resolvent operator maps the normalized vector $\hat{b}$ to the energized output $\hat{q}_{\Psi}$ in the output subspace. 
Finally, the input-output relations are revealed by the SVD of the FI resolvent operator $\boldsymbol{G}$:
\begin{align}
    \boldsymbol{G} (\omega)
    = \sum_j^{N_G} u_{G,j}(\omega)\sigma_{G,j}(\omega){v^*_{G,j}}(\omega) \quad (\sigma_{G,j} \ge \sigma_{G,j+1}> 0),
\end{align}
where $N_G$ is the rank of the matrix $\boldsymbol{G}$.
Using the SVD results, we may recover the $j$-th most energized response mode and the associated forcing mode based on \eqref{eq:basis_in_uspace_f} by selecting $\hat{b}=v_{G,j}$ in \eqref{eq:IO_full} as
\refstepcounter{equation}
\begin{equation*}
  \hat{q}_{G,j} (\omega) = \boldsymbol{G}(\omega)v_{G,j}(\omega) =  u_{G,j}(\omega)\sigma_{G,j}(\omega), 
    \\
    \hat{f}_{G,j} (\omega)= \boldsymbol{T}_{\Phi}(\omega) v_{G,j}(\omega).
    \label{eq:recover}
\tag{$\theequation{\mathit{a}},{\mathit{b}}$}
\end{equation*}
These response mode $\hat{q}_{G,j}$ and forcing mode $\hat{f}_{G,j}$ are what we seek.
Hereafter, we refer to these response and forcing modes extracted via the FI resolvent operator as the FI response mode and FI forcing mode to distinguish from the conventional resolvent modes.
The FI forcing modes $\hat{f}_{G}$ extract the harmonic components of $f'_{\rm{NL}}$, which is essential in sustaining the statistically stationary oscillatory flow motions of interest.
By inserting the definition $\boldsymbol{G}$ and $\boldsymbol{T}_{\Phi}$ into \eqref{eq:recover}, respectively, the FI response and forcing modes are also written as $\hat{q}_{G,j}=\mathbf{\Psi}\mathbf{\Psi}^\dagger\boldsymbol{R}\boldsymbol{T}_{\Phi}v_{G,j}$ and $\hat{f}_{G,j}=\mathbf{\Phi}\mathbf{\Lambda}_{\Phi}^{1/2} v_{G,j}$.
These expressions clearly show that the FI response and forcing modes are expressed as linear combinations of the basis vectors $\mathbf{\Phi}$ of the input subspace and $\mathbf{\Psi}$ of the output subspaces, respectively.
Therefore, the extracted modes always lie within the actual flow field obtained from the numerical simulation.

In practice, when the number of output basis vectors is much smaller than the dimension of the state vector, i.e., $N_{\psi}\ll N_q$, which is typically the case in numerical simulations, it is advantageous to rewrite the relation \eqref{eq:IO_full} by acting $\mathbf{\Psi}^\dagger$ from left as
\begin{align}
    \mathbf{\Lambda}_\Psi^{1/2}(\omega)\hat{a}(\omega) =  \widetilde{\boldsymbol{G}}(\omega)\hat{b}(\omega) \quad\mathrm{where}\quad\widetilde{\boldsymbol{G}}(\omega)= {\mathbf{\Psi}}^\dagger(\omega) \boldsymbol{R}(\omega)\boldsymbol{T}_{\Phi} (\omega)\in \mathbb{C}^{N_\phi\times N_\psi},
    \label{eq:IO_coef}
\end{align}
for the reduced computational cost of the SVD of $\widetilde{\boldsymbol{G}}=\sum_j^{N_G} u'_{G,j}\sigma_{G,j} v_{G,j}$.
We may recover the same response and forcing modes as $\hat{q}_{G,j}=\mathbf{\Psi} u'_{G,j}\sigma_{G,j}$ and $\hat{f}_{G,j}=\boldsymbol{T}_\mathbf{\Phi} v_{G,j}$, respectively.
Furthermore, by taking the inner product of both sides of \eqref{eq:IO_coef}, we obtain 
\begin{equation}
\hat{a}^*(\omega)\mathbf{\Lambda}_{\Psi}(\omega)\hat{a}(\omega)=\hat{b}^*(\omega) \boldsymbol{V}_G^*(\omega) \mathbf{\Sigma}^2_G(\omega) \boldsymbol{V}_G(\omega)\hat{b}(\omega).
\label{recover_output_amp}
\end{equation} 
Here, the original flow's output energy $\mathbf{\Lambda}_{\Psi}$ is decomposed into the singular values of the FI resolvent operator $\mathbf{\Sigma}_{G}$.
Thus, we can examine the most dominant input-output relations by looking at the FI forcing and response modes for the maximum singular value.
In addition, when the FI resolvent operator is a rank-1 matrix ($N_G=1$), the maximum singular value $\sigma_{G,1}$ is expected to match the energy of the output $\hat{q}_\omega$, which will be shown in the results section.

Thus far, we have described the backbone of the FI resolvent analysis framework, discussing how we construct the FI resolvent operator and extract the FI response and forcing modes.
In the subsequent sections, we introduce key properties of the present FI resolvent analysis framework, including the theoretical connection to the input and output basis vectors to modal analysis methods, the physical interpretation of the FI response and forcing mode based on the energy balance in the statistically stationary self-sustained flows in \eqref{eq:energy_balance}, and an alternative view of the FI resolvent operator from the perspective of the state space representation in control theory.

\subsubsection{Use of modal analysis methods for basis vector estimation}\label{sec:modalanalysisforbasis}
We have introduced the orthogonal basis vectors $\mathbf{\Phi}$ and $\mathbf{\Psi}$ for the input and output subspaces, respectively.
A natural choice of these basis vectors is the left singular vectors of the matrices $\widehat{\boldsymbol{F}}$ and $\widehat{\boldsymbol{Q}}$, corresponding to the SPOD modes for the input and output fields, respectively. 
This suggests that some modal analysis techniques are used to estimate the basis vectors of input and output subspaces.
Based on \eqref{eq:basis_in_uspace}, the input and output components in the obtained subspaces are expressed as
\refstepcounter{equation}
\begin{equation*}
    f'_{\Phi}(t) = \int_{-\infty}^{\infty} \mathbf{\Phi}(\omega)\mathbf{\Lambda}^{1/2}_{\Phi} (\omega) \hat{a}(\omega) e^{\mathrm{i}\omega t} d\omega~\in \mathcal{V}_\Phi, 
    \tag{$\theequation\mathit{a}$}
\end{equation*}
\begin{equation*}
    q'_{\Psi}(t) =\int_{-\infty}^{\infty} \mathbf{\Psi}(\omega)\mathbf{\Lambda}^{1/2}_{\Psi} (\omega) \hat{b}(\omega )e^{\mathrm{i}\omega t} d\omega ~\in \mathcal{V}_{\Psi}.
    \tag{$\theequation\mathit{b}$}
\end{equation*}
This is a general form of the standard modal decompositions in which the spatial coherence ($\mathbf{\Phi},\mathbf{\Psi}$) and temporal coherence (harmonic component $e^{\mathrm{i}\omega t}$) are considered separately \citep{Taira2017-kn}.
This form can cover cases where the non-orthogonal basis vectors are considered as discussed in Appendix \ref{app:rankdecom}.
Therefore, modal analysis methods, such as SPOD (orthogonal) and DMD (typically non-orthogonal), are applicable for estimating the bases of the input and output subspaces.
Regarding the DMD modes, those whose eigenvalues lie on the unit circle serve as candidate basis vectors for the Fourier-based formulation considered here.

The data-driven resolvent analysis uses the DMD modes of the output $q'$ as the basis vectors, not only for the output subspace but also for the input subspace \citep{Herrmann2021-fk}.
In this sense, the present FI resolvent analysis, which considers different bases for the input and output subspaces, offers a more general formulation.

\subsubsection{Time-averaged energy balance for FI response and forcing modes}
As discussed earlier in \eqref{eq:timeave_evolution_of_Eq'}, the time-averaged energy balance between the energy amplification/attenuation balance between the linear mechanism $\overline{\langle q', \boldsymbol{L}q'\rangle_W}$ and the nonlinear mechanism $\overline{\langle q', f'_{\rm NL}\rangle_W}$ is the key property of the self-sustaining mechanisms. 
Using the identified response and forcing modes, we express $q'$ and $f'_{\rm{NL}}$ as their linear combinations, respectively written as
\refstepcounter{equation}
\begin{equation*}
    q' = \int_{0}^{\infty}\left( \sum_{j=1}^{N_G} \hat{q}_{G,j}(\omega)\right)e^{i\omega_m t} d\omega_m +\mathrm{c}.\mathrm{c}.,
    \quad
    f'_{\rm{NL}} = \int_{0}^\infty \left(\sum_{k=1}^{N_G} \hat{f}_{G,k}(\omega)\right)e^{\mathrm{i}\omega_n t} d\omega_n+\mathrm{c}.\mathrm{c}.,
\tag{$\theequation{\mathit{a},\mathit{b}}$}\label{eq:recov_by_opt}
\end{equation*}
where $\rm c.c.$ denotes the complex conjugate.
By inserting these relations into \eqref{eq:timeave_evolution_of_Eq'} and using the orthogonality of the Fourier basis $\overline{\langle e^{i\omega_m, t}, e^{i\omega_n t}\rangle}=\delta_D(\omega_m-\omega_n)$, with $\delta_D$ being the Dirac delta, we obtain the following relation at frequency $\omega$: 
\begin{align}
    0
    =
    \sum_{j=1}^{N_G}\sum_{k=1}^{N_G} 
    \left( 
    { \hat{q}_{G,j} }^{*}(\omega) \boldsymbol{W}\boldsymbol{L}\hat{q}_{G,k} (\omega)
    + 
    { \hat{q}_{G,j}}^{*}(\omega)  \boldsymbol{W}\hat{f}_{G,k} (\omega)
    \right)
    +\mathrm{c}.\mathrm{c}..
    \label{eq:energy_balance}
\end{align}

In particular, when the rank-1 approximation of the FI resolvent operator is valid (i.e.,~$\sigma_{G,1}\gg\sigma_{G,2}$), the above relation \eqref{eq:energy_balance} may be approximated by setting $N_G=1$ as
\begin{align}
     0
    &\approx
    \left( 
    { \hat{q}_{G,1}}^{*}(\omega) \boldsymbol{W}\boldsymbol{L}\hat{q}_{G,1} (\omega)
    +
    { \hat{q}_{G,1}}^{*} (\omega) \boldsymbol{W}\hat{f}_{G,1} (\omega)
    \right)
    +\mathrm{c}.\mathrm{c}..\nonumber\\
    &= \mathrm{Re}\left( 
    { \hat{q}_{G,1}}^{*} (\omega)\boldsymbol{W}\boldsymbol{L}\hat{q}_{G,1} (\omega)
    \right)
    +
    \mathrm{Re}\left( 
    { \hat{q}_{G,1}}^{*} (\omega) \boldsymbol{W}\hat{f}_{G,1} (\omega)
    \right).
    \label{eq:energy_balance_lowrank}
\end{align}
Note that, before using \eqref{eq:energy_balance_lowrank}, the validity of the rank-1 approximation should be assessed by looking at the gap between the leading and secondary singular values.
As shown later, this approximation is valid for the cases considered in this paper.
When this approximation is valid, the dominant nonlinear energy amplification/attenuation mechanisms are expressed by the second term in \eqref{eq:energy_balance_lowrank}.
Based on this term, we can locate the active spatial region where nonlinear energy amplification or attenuation is prevalent by visualizing the Hadamard (element-wise) product $\odot$ of the leading response $\hat{q}_{G,1}$ and forcing $\hat{f}_{G,1}$, written as 
\begin{equation}
  \mathcal{T}(\omega)=  \mathrm{Re} \left[(\boldsymbol{Z}_W\hat{q}_{G,1}(\omega))^*\odot (\boldsymbol{Z}_W\hat{f}_{G,1}(\omega))\right] \in\mathbb{R}^{N_q}.
    \label{eq:NLenergyTransferMap}
\end{equation}
The summation over the elements of vector $\mathcal{T}$ reduces to the second term in  \eqref{eq:energy_balance_lowrank}.
Following \citet{Jin2021-ge}, we call $\mathcal{T}$ the nonlinear energy transfer map, because the inner product $\langle q', f'_{\rm{NL}}\rangle$ contains the triadic interaction of three waves that is responsible for the nonlinear inter-scale energy transfer.
Since we directly compute $f'_{\rm{NL}}$ now, all possible combinations of the wavenumber triads (in the range of the scales resolved by the numerical simulation) are contained in $\langle q', f'_{\rm{NL}}\rangle$.
Specifying the wavenumber triads in computing $f'_{\rm{NL}}$ is possible and may provide an opportunity to study inter-scale interactions and energy transfer among the modes (e.g., \citealt{Yeung2026-qb,Nakamura2026-ni}).

\subsubsection{FI resolvent analysis in terms of state-space representation}
The derived FI resolvent operator $\boldsymbol{G}(\omega) =  \boldsymbol{P}_{\Psi}(\omega) \boldsymbol{R}(\omega) \boldsymbol{T}_\Phi(\omega)$ in \eqref{eq:FIresolventoperator} is the transfer function of the following state-space system:
\begin{align}
    \left\{
    \begin{aligned}
    \frac{\partial q'}{\partial t} &= \boldsymbol{L}q' + f'_{\Phi_\omega}\\
    f'_{\Phi_\omega}& = B_{\Phi_\omega}b' \in\mathcal{V}_{\Phi_\omega}\\
    q'_{\Psi_\omega}&= C_{\Psi_\omega} q' \in\mathcal{V}_{\Psi_\omega},
    \end{aligned}
    \right.
    \label{eq:windowedsystem}
\end{align}
where $B_{\Phi_\omega}$ and $C_{\Psi_\omega}$ are the linear operators respectively defined as
\refstepcounter{equation}
\begin{equation*}
B_{\Phi_\omega}(\omega):= \mathcal{F}^{-1}\circ \boldsymbol{T}_{\Phi}(\omega)\circ\mathcal{P}(\omega)\circ \mathcal{F},
\quad 
C_{\Psi_\omega}(\omega):= \mathcal{F}^{-1}\circ \boldsymbol{P}_{\Psi}(\omega)\circ \mathcal{P}(\omega)\circ \mathcal{F}.
\tag{$\theequation{\mathit{a},\mathit{b}}$}
\end{equation*}
Here, $\mathcal{F}$ and $\mathcal{F}^{-1}$ are the Fourier transform operator and its inverse, $\mathcal{P}(\omega)$ is the projection operator onto a specific frequency $\omega$ of interest, and the operator $\circ$ indicates the composition.
The input noise is defined as $b':= \mathcal{F}^{-1}\hat{b}$. With this definition, the Fourier transform $\mathcal{F}$ of  \eqref{eq:windowedsystem} yields $\hat{q}_\omega = \boldsymbol{G}\hat{b}_\omega$ \eqref{eq:IO_full}.
These linear operators $B_{\Phi_\omega}$ and $C_{\Psi_\omega}$ act as spatiotemporal filters that express the component with the spatial coherence (prescribed by the bases of input subspace $\mathbf{\Phi}$ and output subspace $\mathbf{\Psi}$) and temporal dynamics (governed by $e^{\mathrm{i}\omega t}$).
As such, the present FI resolvent analysis is viewed as an extension of the input-output analysis \citep{Jovanovic2005-rq,Jeun2016-gi}. 
The present method constructs the windowing matrix from the numerical simulation data. 
This interpretation may provide opportunities to apply well-established control theory, such as observability, controllability, and balanced truncation, to the FI resolvent analysis.

The noise $b'$ is the white noise vector, which is uncorrelated in terms of time, space, and variables such that  $E[b'_i(t)b'_j(\tau)]=\delta_{ij}\delta_D(t-\tau)$.
The energy of the noise $b'$ is distributed uniformly across the all the frequency $\omega$ as $\|\hat{b}\|_2=1$, and hence we may ensure that all the characteristics of nonlinear forcing at frequency $\omega$ arises only from the input filter $B_{\Phi_\omega}$ and that the energy gain by transformation $\boldsymbol{G}$ (i.e., the singular values of $\boldsymbol{G}$) correspond to the amplitudes of the response mode.
In light of the discussion of \citet{Towne2018-zc},  the conventional resolvent analysis corresponds to the case of $B_{\Psi}=C_{\Phi}=\mathcal{I}$(identity map).

Let us mention in passing that it is possible to introduce a general integral transformation, such as the Laplace transformation, the wavelet transformation, instead of the Fourier transform of system \eqref{eq:windowedsystem}, as considered in the context of the resolvent analysis \citep{Yeh2020-uu, Ballouz2024-wk, Lopez-Doriga2024-uk}.
The choice of kernel function in the integral transform directly characterizes the temporal structure of the FI response and forcing modes. 
For instance, the Fourier kernel considered here yields the harmonic modes.

\section{Numerical Algorithms}\label{sec:numalg}
The numerical algorithms employed to compute the FI resolvent operator and to obtain the response and forcing modes are described here.
Here, we present two types of formulations depending on the computational approach for the linear operator $\boldsymbol{L}$, while the same algorithm is applied to estimate $\boldsymbol{T}_{\Phi}$ and $\boldsymbol{P}_{\Psi}$ used to construct the FI resolvent operator.
One is the fully data-driven approach, in which the linear operator $\boldsymbol{L}$ is estimated in a data-driven manner utilizing the snapshots of nonlinear forcing $f'_{\rm{NL}}$. 
We mainly use this fully data-driven approach in this paper.
The other is the semi-data-driven approach, in which the linear operator $\boldsymbol{L}$ is obtained via conventional algorithms developed in the context of global stability analysis (e.g., \citealt{Chiba1998,Mettot2014-sf}, and see also the reviews by \citealt{Theofilis2011-bi} and \citealt{Rolandi2024-jr} for details). 
We denote the FI resolvent operator computed by the fully data-driven approach as $\boldsymbol{G}_{\rm{FDD}}$ and that computed by the semi-data-driven approach as $\boldsymbol{G}_{\rm{SDD}}$.

\subsection[dummyforindex]{Estimations of $\boldsymbol{T}_{\Phi}$ and $\boldsymbol{P}_\Phi$ using SPOD}{\label{sec:estTP_via_SPOD}}
We use the SPOD method to estimate the bases of input and output subspaces.
To obtain the SPOD modes, the CSD tensors of input $\hat{f}_{\rm NL}$ and output $\hat{q}$ are first estimated as
\refstepcounter{equation}
\begin{equation*}
\boldsymbol{S}_{\rm{ff}}= \kappa\widehat{\boldsymbol{F}} \widehat{\boldsymbol{F}}^*,
\qquad
\boldsymbol{S}_{\rm{qq}}= \kappa\widehat{\boldsymbol{Q}} \widehat{\boldsymbol{Q}}^*, 
\tag{$\theequation{\mathit{a},\mathit{b}}$}\label{eq:est_CSDtensor}
\end{equation*}
where $\kappa = \Delta t/(\alpha N_p)$, $\Delta t$ is the sampling time step, and $\alpha$ is a correction factor of a window function used in the discrete Fourier transform for mitigating spectral leakages.
The SPOD modes are obtained by the eigenvalue decomposition of the CSD tensors:
\refstepcounter{equation}
\begin{equation*}
\boldsymbol{S}_{\rm{ff}}\boldsymbol{W}= \boldsymbol{\Phi}_{\rm{SPOD}}\boldsymbol{\Lambda}_{ \Phi_{\rm{SPOD}}} \boldsymbol{\Phi}_{\rm{SPOD}}^*,
\qquad  
\boldsymbol{S}_{\rm{qq}}\boldsymbol{W}= \boldsymbol{\Psi}_{\rm{SPOD}}\boldsymbol{\Lambda}_{ \Psi_{\rm{SPOD}}} \boldsymbol{\Psi}_{\rm{SPOD}}^*, 
\tag{$\theequation{\mathit{a},\mathit{b}}$}\label{eq:EVDofCSDtensors}
\end{equation*}
where $\boldsymbol{\Lambda}_{ \Phi_{\rm{SPOD}}}$ and $\boldsymbol{\Lambda}_{ \Psi_{\rm{SPOD}}}\in \mathbb{R}^{N_p\times N_p}$ are the eigenvalues of $\boldsymbol{S}_{\rm{ff}}$ and $\boldsymbol{S}_{\rm{qq}}$. 
Also, $\boldsymbol{\Phi}_{\rm{SPOD}}$ and $\boldsymbol{\Psi}_{\rm{SPOD}}\in \mathbb{C}^{N_q\times N_p}$ are the corresponding $N_p$ eigenvectors, respectively. 
The SPOD modes satisfy the weighted orthonormal conditions of $\boldsymbol{\Phi}_{\rm SPOD}^*\boldsymbol{W}\boldsymbol{\Phi}_{\rm SPOD}=\boldsymbol{I}_{\rm N_p}$ and $\boldsymbol{\Psi}_{\rm SPOD}^*\boldsymbol{W}\boldsymbol{\Psi}_{\rm SPOD}=\boldsymbol{I}_{\rm N_p}$, respectively.
Using these SPOD modes, the matrices $\boldsymbol{T}_{\Phi}$ and $\boldsymbol{P}_\Phi$ are expressed as
\stepcounter{equation}
\begin{equation*}
    \boldsymbol{T}_{\Phi} = \mathbf{\Phi}_{\rm{SPOD}}\mathbf{\Lambda}_{\Phi_{\rm{SPOD}}}^{1/2},
    \qquad 
    \boldsymbol{P}_{\Psi} = \mathbf{\Psi}_{\rm{SPOD}}\mathbf{\Psi}_{\rm{SPOD}}^\dagger.
\tag{$\theequation{\mathit{a},\mathit{b}}$}\label{eq:SPODbasedTP}
\end{equation*}
When the SPOD modes are used for the output bases, the output amplitude $\mathbf{\Lambda}_{\Psi}^{1/2}$ becomes the square root of the SPOD gains.
Hence, we will compare the gain of the FI resolvent operator and the square root of the output SPOD gain in the result section.

\subsection[dummyforindex]{Fully data-driven approach with estimated $\boldsymbol{L}$}\label{sec:fully-data-driven}
As mentioned in the introduction, another objective of the present study is to estimate the FI resolvent operator $\boldsymbol{G}$ in a data-driven manner, while reducing the dimensionality of the large-scale flow dataset.
Here, we present a complete data-driven formulation, in which the linear operator $\boldsymbol{L}$ is estimated from snapshot data of both output $q'$ and input $f'_{\rm{NL}}$.
To do so, we use the governing equation of fluctuations \eqref{eq:NLpert}, which is rearranged as
\begin{align}
    \frac{\partial q'}{\partial t} - f'_{\rm NL} = \boldsymbol{L}q'.
    \label{eq:ttt3}
\end{align}
Alongside, we define the snapshot matrices of output and input as
\begin{align}
  \boldsymbol{F} &= \left[
    \begin{matrix}
        {f}'_{\mathrm{NL},1} & {f}'_{\mathrm{NL},2}& \cdots & {f}'_{\mathrm{NL},M}
    \end{matrix}
    \right]\in \mathbb{R}^{N_q\times M},\\
    \boldsymbol{Q} &= \left[
    \begin{matrix}
        {q}'_{1} & {q}'_{2} & \cdots & {q}'_{t_M}
    \end{matrix}
    \right] \in \mathbb{R} ^{N_q\times M},
\end{align}
respectively, and we also introduce the matrix of the time derivatives of output snapshots as
\begin{align}
    \dot{\boldsymbol{Q}} &= \left[
    \begin{matrix}
        {(dq'/dt)}_1 & {(dq'/dt)}_2 & \cdots & {(dq'/dt)}_M
    \end{matrix}
    \right] \in \mathbb{R} ^{N_q\times M}.
\end{align}
Since \eqref{eq:ttt3} holds at every snapshot, we may obtain the following relation
\begin{align}
    \dot{\boldsymbol{Q}} - \boldsymbol{F} = \boldsymbol{L}\boldsymbol{Q}.
    \label{eq:matrixgoveningeq}
\end{align}
Based on \eqref{eq:matrixgoveningeq}, the linear time invariant operator $\boldsymbol{L}$ can be estimated by solving the following minimisation problem:
\begin{align}
\boldsymbol{L}\approx \boldsymbol{L}_{\rm{data}} = \arg\min_{\boldsymbol{L}} \|  \dot{\boldsymbol{Q}}-\boldsymbol{F}-\boldsymbol{L}\boldsymbol{Q} \|_F,
\label{eq:minimizationproblem}
\end{align}
where $\|\cdot\|_F$ denotes the Frobenius norm. 
The solution to this problem is given as
\begin{align}
    \boldsymbol{L}_{\rm{data}}= ( \dot{\boldsymbol{Q}}- \boldsymbol{F}) \boldsymbol{Q}^+ ,
    \label{eq:estimate_L}
\end{align}
where $\boldsymbol{Q}^{+}:= \boldsymbol{V}_Q \boldsymbol{\Sigma}_Q^{-1}\boldsymbol{U}_Q^*$  is the pseudo-inverse of $\boldsymbol{Q}$, obtained using the SVD $\boldsymbol{Q}=\boldsymbol{U}_Q\boldsymbol{\Sigma}_Q\boldsymbol{V}^*_Q$.

In practice, to obtain $\boldsymbol{L}_{\rm{data}}$, a similar algorithm used in the projected DMD \citep{Schmid2010-gx}, or its variants, such as exact DMD \citep{Tu2014-lu} and parallel algorithms for a large-scaled data \citep{Asada2025-ib} are applicable, by dealing with the left-hand-side term $\dot{\boldsymbol{Q}}-\boldsymbol{F}$ in \eqref{eq:matrixgoveningeq} in the same way as the snapshot matrix of next time step in the projected DMD.
Hence, as in the projected DMD, equation \eqref{eq:matrixgoveningeq} is projected onto the first $N_{\mathrm{POD}}$ leading POD modes of $\boldsymbol{Q}$, denoted as $\boldsymbol{U}_{Q}^{N_{\mathrm{POD}}}$, which are the left singular vectors corresponding to the first $N_{\mathrm{POD}}$ leading singular values $\mathbf{\Sigma}_{Q}^{N_{\rm{POD}}}$ of $\boldsymbol{Q}$. 
The reduced-sized linear operator on the projected equation is estimated as 
\begin{align}
   \widetilde{\boldsymbol{L}}_{\rm{data}}=({\boldsymbol{U}_Q^{N_{\rm{POD}}}})^*(\dot{\boldsymbol{Q}}  - \boldsymbol{F}) \boldsymbol{V}_Q^{N_{\rm{POD}}}{(\boldsymbol{\Sigma}_Q^{N_{\rm{POD}}}})^{-1} \in \mathbb{R}^{N_{\rm{POD}}\times N_{\rm{POD}}},
\end{align}
where $\boldsymbol{V}_{Q}^{\rm{POD}}$ collects the right singular vectors of $\boldsymbol{Q}$ corresponding to $\mathbf{\Sigma}_{Q}^{\rm{POD}}$.
The original-sized linear operator is recovered as
\begin{equation}
\boldsymbol{L}_{\rm{data}}=\boldsymbol{U}_Q^{N_{\rm{pod}}}\widetilde{\boldsymbol{L}}_{\rm{data}}(\boldsymbol{U}_Q^{N_{\rm{pod}}})^*.
\label{eq:ROM_L}
\end{equation}
Finally, inserting $\boldsymbol{L}=\boldsymbol{L}_{\rm{data}}$ into \eqref{eq:IO_coef}, the fully data-driven FI resolvent operator $\widetilde{\boldsymbol{G}}_{\rm{FDD}}$ is rewritten as 
\begin{align}
    \widetilde{\boldsymbol{G}}_{\rm{FDD}} 
    &= \boldsymbol{\Psi}^\dagger (\mathrm{i}\omega \boldsymbol{I}-\boldsymbol{L}_{\rm{data}}) \boldsymbol{\Phi}\boldsymbol{\Lambda}_{\Phi}
\nonumber\\
    &=\boldsymbol{\Psi}^\dagger
    \boldsymbol{U}^{N_{\rm{pod}}}_Q (\mathrm{i}\omega \boldsymbol{I}_{N_{\mathrm{POD}}} - \widetilde{\boldsymbol{L}}_{\rm{data}})^{-1} {(\boldsymbol{U}^{N_{\rm{pod}}}_Q)}^*\boldsymbol{\Phi}\boldsymbol{\Lambda}_{\Phi}
    ,
    \label{eq:fully data-driven-G}
\end{align}
where $\boldsymbol{I}_{N_{\rm{POD}}}$ is the $N_{\rm POD}\times N_{\rm POD}$  identity matrix.
After the SVD of $\widetilde{\boldsymbol{G}}_{\rm{FDD}}$, the FI response and forcing modes are recovered in the same way as in \eqref{eq:recover}.

It is also worth discussing that to obtain the linear time invariant operator $\boldsymbol{L}$, although one may want to use the dynamic mode decomposition (DMD) \citep{Schmid2010-gx} as in the data-driven resolvent analysis of linear flows \citep{Herrmann2021-fk}, this approach does not work in this situation where the nonlinear forcing terms $f'_{\rm NL}$ have finite amplitude. 
 This is because the DMD attempts to find a time evolution operator $\boldsymbol{A}$ such that $\boldsymbol{Q}'\approx \boldsymbol{A}\boldsymbol{Q} \approx \int_{\Delta t} \boldsymbol{L}\boldsymbol{Q}+\boldsymbol{F} dt$ ($\boldsymbol{Q}'$ is the matrix collecting the snapshots of the next time steps with time increment $\Delta t$), which includes the effects of nonlinear forcing terms $\boldsymbol{F}$.
Although the DMD is favorable for approximating a good time-evolution operator for nonlinear dynamics, it can face difficulty in finding $\boldsymbol{L}$, which will be shown for the Stuart-Landau oscillator in \S\ref{sec:3.1}.

\subsection{Semi-data-driven approach}\label{sec:semi-data-driven}
Next, let us consider the situation where we have the linear operator $\boldsymbol{L}$ obtained using one of several numerical algorithms developed for global stability analysis \citep{Theofilis2011-bi}.
The algorithm for the input-output analysis based on the semi-data-driven $\boldsymbol{G}_{\rm{SDD}}$ is shown in algorithm \ref{alg:opbased_resolvent} below.
The advantage of this algorithm is that we may avoid the direct inverse operation to have the resolvent operator $\boldsymbol{R}$ by taking its action on the input $\boldsymbol{T}_{\Phi}$ (in line 1) using a linear solver, which is cost-efficient and stable.
The present algorithm is similar to the randomized resolvent analysis \citep{Ribeiro2020-gh}, and thus achieves comparably reduced computational cost.

\begin{algorithm}[H]
\caption{Semi-data-driven forcing-informed resolvent analysis}
\centering
\begin{algorithmic}[1]
\renewcommand{\algorithmicrequire}{\textbf{Input:}}
\renewcommand{\algorithmicensure}{\textbf{Output:}}
\Require Frequency $\omega$, linear operator $\boldsymbol{L}$, and matrices $\boldsymbol{T}_{\Phi}$ and $\mathbf{\Phi}^\dagger$
\State $\boldsymbol{G}_{\rm{SDD}}\gets [\mathrm{i}\omega \boldsymbol{I}_{\rm N_q}-\boldsymbol{L}] \backslash \boldsymbol{T}_{\Phi}$  \Comment{Solve linear system for $\boldsymbol{G}_{\rm{SDD}}$ } 
\State $\widetilde{\boldsymbol{G}}_{\rm{SDD}}\gets\boldsymbol{\Psi}^\dagger \boldsymbol{G}_{\rm{SDD}}$ \Comment{Projection onto the output (coefficient) space estimated from data}
\State $[\boldsymbol{U}_{\widetilde{G}},\boldsymbol{\Sigma}_{\widetilde{G}},\boldsymbol{V}_{\widetilde{G}}]\gets \rm{svd}({\widetilde{\boldsymbol{G}}_{\rm{SDD}}})$ \Comment{SVD of reduced-sized $\widetilde{\boldsymbol{G}}_{\rm{SDD}}$ }
 \ForAll {$j = 1$ to $N_G$} 
\State $[\hat{q}_{G,j},\hat{f}_{G,j}] \gets [\boldsymbol{\Psi}u_{\widetilde{G},j}\sigma_{\widetilde{G},j},\boldsymbol{\Phi}\mathbf{\Lambda}_{\Phi}^{1/2}v_{\widetilde{G},j}]$ 
 \EndFor \Comment{Recover $j-$th leading FI response/forcing modes}
\Ensure Pairs of FI forcing and response modes
\end{algorithmic}
\label{alg:opbased_resolvent}
\end{algorithm}
For the linear solver on line 1, a direct sparse LU method is employed in this study. 
In the randomized resolvent analysis \citep{Ribeiro2020-gh}, the sketch of the resolvent $\boldsymbol{R}$ is first obtained using the random test matrix $\boldsymbol{\Omega}$ by solving the linear system $\boldsymbol{R}^{-1}\boldsymbol{S}_R=\boldsymbol{\Omega}$ for the sketch $\boldsymbol{S}_R$.
By this operation, we obtain the action of the resolvent operator $\boldsymbol{R}$.
In the present method, the first line corresponds to this process; the action of $\boldsymbol{R}$ is taken on $\boldsymbol{T}_{\Phi}$.
\citet{Ribeiro2020-gh} also introduced the physics-informed random test matrix $\mathbf{\Omega}_{\rm{physics}}$ instead the use of random matrix $\mathbf{\Omega}$. 
The entries of $\mathbf{\Omega}_{\rm{physics}}$ are scaled by the velocity gradient of the base flow, magnifying the non-normality associated with the shear.
It was reported that introducing a physics-informed random test matrix yields more accurate resolvent modes with fewer column vectors of the test matrix.
In this way, $\boldsymbol{T}_{\Phi}$ can be viewed as a physics-informed test matrix that yields the response excited by the nonlinear forcing.
On the other hand, the projection onto the output coefficient space in the second line corresponds to the projection onto the column space of the sketch $\boldsymbol{S}_R$ in the randomized resolvent analysis.

\section{Summary of the FI resolvent analysis}\label{sec:summary}
Before the results section, the overall workflow of the present FI resolvent analysis is summarized below again, with reference to Fig.~\ref{fig:summary}.

     In step 1, we begin by collecting the output snapshots $\boldsymbol{Q}$ and computing the nonlinear forcing snapshots $\boldsymbol{F}$, which are necessary in self-sustaining mechanisms as discussed in \S\ref{sec:2.1}. 
    For compressible NSs considered here, the nonlinear forcing is given in Appendix~\ref{app:forcingterm}.
    
     In step 2, the basis vectors $\mathbf{\Psi}$ and $\mathbf{\Phi}$ of input and output subspaces are estimated from their data, respectively.
    Specifically, modal analysis methods can be used to estimate their basis vectors, as discussed in \S\ref{sec:modalanalysisforbasis}. 
    This paper uses the SPOD modes as a natural choice for the Fourier-transform-based formulation of the resolvent operator, as detailed in \S\ref{sec:estTP_via_SPOD}.
    When choosing the fully data-driven implementation, the linear operator $\boldsymbol{L}$ can also be estimated by leveraging the forcing snapshots, as explained in \S\ref{sec:fully-data-driven}.
    
     In step 3, using the obtained nonlinear forcing information contained in $\boldsymbol{T}_{\Phi}$ and projection operator onto the output subspace $\boldsymbol{P}_{\Psi}$, the FI resolvent operator $\boldsymbol{G}$ is constructed as in \eqref{eq:FIresolventoperator}. 
    The computational cost of $\boldsymbol{G}$ can be reduced using \eqref{eq:fully data-driven-G} for the fully data-driven approach in \S\ref{sec:fully-data-driven} or Algorithm~\ref{alg:opbased_resolvent} for the semi-data-driven approach as in \S\ref{sec:semi-data-driven}.
    
     In step 4, the FI forcing and response modes are recovered via the SVD of the FI resolvent operator, as given by \eqref{eq:recover}. 
    This relation \eqref{eq:recover} ensures that these modes lie in the estimated input $\mathcal{V}_{\Phi}$ and output subspaces $\mathcal{V}_{\Psi}$, respectively. 
    In other words, the FI resolvent analysis identifies the FI forcing and response modes that exist in the actual flow field. 
    This property holds even at frequencies where nonlinear amplification mechanisms dominate and thus are not revealed by conventional resolvent analysis, as indicated by the modes at $2\omega_1$ for the 2D cylinder example in  Fig.~\ref{fig:summary}. 
    The details for the results of this 2D cylinder case are given in the subsequent section \ref{sec:res_2Dcylinder}.

\section{Results and discussion}\label{sec:results}

\subsection{Stuart-Landau oscillator}\label{sec:3.1}
As a proof of concept for the proposed method, let us consider a model dynamical system of a third-order Stuart-Landau oscillator, which models dynamics near the supercritical Hopf bifurcation.
The Hopf bifurcation can be found in several flow systems, including two-dimensional cylinder flow \citep{Sipp2007-fw}  and transonic airfoil buffeting flow \citep{Crouch2007-wr}.
This model can be expressed as
\begin{align}
    \frac{d}{dt}\left[
    \begin{matrix}
    \alpha\\\beta
    \end{matrix}\right] 
    =
    \left[
    \begin{matrix}
    \lambda_L & \omega_L \\
    -\omega_L & \lambda_L
    \end{matrix}\right]\left[
    \begin{matrix}
    \alpha\\\beta
    \end{matrix}\right] 
    - 
    (\alpha^2+\beta^2)
    \left[\begin{matrix}
    \alpha-c\beta \\
    c\alpha+\beta
    \end{matrix}\right] ,
    \label{eq:SL3rd_mat}
\end{align}
for a state vector of $q'=[\alpha,\beta]\in \mathbb{R}^2$ and parameters $\lambda_L, \omega_L,c\in\mathbb{R}$.
The nonlinear fluctuation equations \eqref{eq:NLpert} can be derived not only for the Navier--Stokes equations but also for a general nonlinear governing equation.
In \eqref{eq:SL3rd_mat}, the first term in the right-hand side corresponds to the linear term $\boldsymbol{L}q'$ while the second term to the nonlinear forcing term $f'_{\rm NL}$.

\begin{figure}
    \centering
    \includegraphics[width=0.9\linewidth]{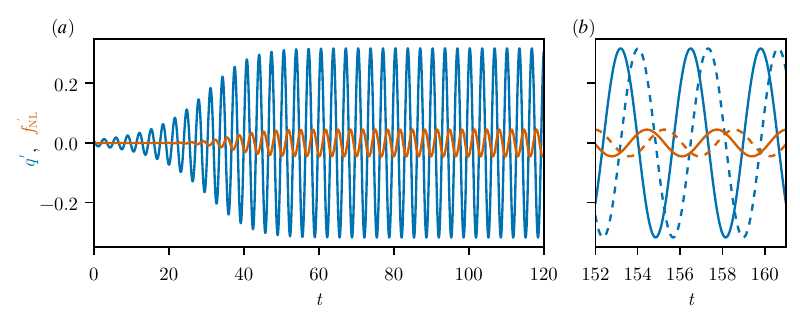}
    \caption{Time trajectories of the state vector $q'$ (blue) and the nonlinear forcing vector $f'_{\rm NL}$ (orange) of the Stuart--Landau oscillator. (\textit{a}) The first components $q'_1$ and $f'_{\rm NL,1}$ from the initial condition to the limit cycle oscillation (LCO) at $0\le t\le 120$. (\textit{b}) All components at $152\le t\le 161$ in the LCO domain. Solid lines, the first components of $q'$ and $f'_{\rm{NL}}$; dashed lines, the second components.}
    \label{fig:SL3rd_timehist}
\end{figure}

Here, we set the parameters as $\lambda_L =0.10, \omega_L=2.0$, and $c=1.0$ and consider an initial condition of $[\alpha(0),\beta(0)]^t= 0.01[1/\sqrt{2},1/\sqrt{2}]^t$.
With the parameters and the initial condition, the solution $q'$ and the nonlinear forcing term $f'_{\rm{NL}}$ are as shown in Fig.~\ref{fig:SL3rd_timehist}(\textit{a}). 
The amplitude of the solution grows exponentially, and subsequently it reaches the limit cycle around $t=60$.
As shown in Fig.~\ref{fig:SL3rd_timehist}(\textit{b}), the nonlinear forcing term $f'_{\mathrm{NL}}$ also varies in time periodically, similar to the output $q'$, with a fixed phase difference with $q'$.
In the exponentially growing region, the solution oscillates at a frequency of $\omega=\omega_L=2.0$. On the other hand, the frequency of the limit-cycle oscillation (LCO) is analytically given as $\omega_{\rm{NL}}=\omega_L-c\lambda_L=1.9$. 
For this Stuart-Landau model, let us assess the performance of the proposed FI resolvent framework by checking if (1) the linear time-invariant matrix $\boldsymbol{L}=[ \lambda_L ~\omega_L; -\omega_L~\lambda_L]$ is correctly estimated, (2) the proposed FI resolvent operator $\boldsymbol{G}$ can predict the nonlinear frequency $\omega_{\rm NL}=1.9$, and (3) the energy balance between linear and nonlinear amplification/attenuation mechanisms as in \eqref{eq:timeave_evolution_of_Eq'} holds.

\begin{figure}
    \centering
    \includegraphics[width=1\linewidth]{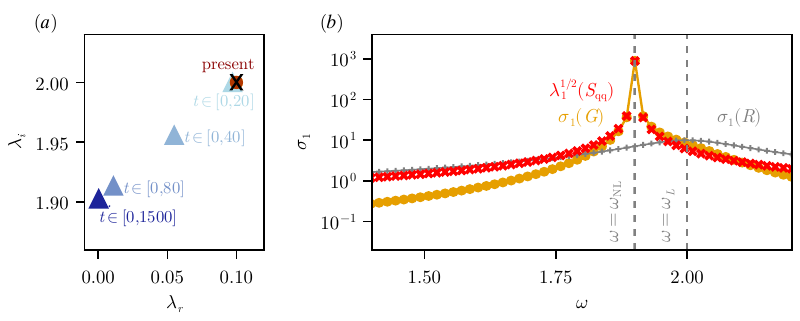}
    \caption{Results for the Stuart-Landau oscillator. (\textit{a}) Eigenvalue $\lambda_r+\mathrm{i}\lambda_i$ of $\boldsymbol{L}_{\mathrm{data}}$ obtained by the proposed method (red circle) and by DMD (blue triangle) and the true eigenvalue $(\lambda_r,\lambda_i)=(0.1,2.0)$ (black cross).
    The red circles collapse onto the true eigenvalue regardless of the time ranges.
    (\textit{b}) Leading singular value $\sigma_1$ of proposed fully data-driven FI resolvent operator $\boldsymbol{G}_{\rm{FDD}}$ (orange circle) and conventional resolvent operator $\boldsymbol{R}$ (gray plus), and square roots of the leading SPOD gains (red cross).}
    \label{fig:eval_gain_SL3rd}
\end{figure}

For the first point (1), we consider the four datasets with different time ranges $t\in[t_s,t_e]$, specifically, $[t_s,t_e]=[0,20]$ (linear growth domain), $[0,40]$, $[0,80]$ (linear growth and LOC domains are mixed to the same degree),  and $[0,1500]$ (almost covered by the LCO), with a sufficiently small sampling interval $\Delta t=1.0\times10^{-2}$. 
In this test case, time derivatives $\dot{\boldsymbol{Q}}$ are evaluated using the second-order central differencing.
For comparison, the DMD is used to estimate $\boldsymbol{L}$. 
With the DMD, $\boldsymbol{L}$ can be recovered as $\boldsymbol{V}_{\rm DMD}\log(\boldsymbol{\Lambda}_{\rm DMD})/\Delta t \boldsymbol{V}_{\rm DMD}^{-1}$ where $\boldsymbol{V}_{\rm DMD}$ and $\boldsymbol{\Lambda}_{\rm DMD}$ are the eigenvectors (DMD modes) and eigenvalues of the linear time-evolution matrix.
In figure \ref{fig:eval_gain_SL3rd}(\textit{a}), the eigenvalues of the estimated $\boldsymbol{L}_{\rm{data}}$ by the proposed data-driven approach and the DMD are shown.
Since the theoretical eigenvalues of $\boldsymbol{L}$ are $\lambda_L\pm \mathrm{i}\omega_L$, the real part and imaginary part of the estimated eigenvalues should correspond to the diagonal ($\lambda_L$) and off-diagonal elements ($\omega_L$) of $\boldsymbol{L}$, respectively.
Thus, we can see if the linear operator is predicted or not by visualizing the eigenvalue (only one from the complex conjugate pair) as in Fig.~\ref{fig:eval_gain_SL3rd}(\textit{a}).
The present method well predicts $\boldsymbol{L}$ robustly over the time range, as shown by the red points collapsing onto the reference black cross marker.
In contrast, DMD is unable to predict $\boldsymbol{L}$ when the LCO solution accounts for most of the time range.
This is because the term $\alpha^2+\beta^2$ becomes constant in the LCO region of a constant amplitude, which simplifies the Stuart-Landau model \eqref{eq:SL3rd_mat} to a linear model, written as
\begin{align}
    \frac{d}{dt}\left[
    \begin{matrix}
    \alpha\\\beta
    \end{matrix}\right] 
    =
    \underset{\boldsymbol{L}'}{
    \underbrace{
    \left[
    \begin{matrix}
    (\lambda_L-d) & \omega_L-cd \\
    -(\omega_L-cd) & \lambda_L-d
    \end{matrix}\right]}}
    \left[
    \begin{matrix}
    \alpha\\\beta
    \end{matrix}\right]
    \label{eq:SL3rd_mat_inLCO}
\end{align}
where $d=\alpha^2+\beta^2$ is the constant amplitude.
The DMD predicts $\boldsymbol{L}'$ that includes the forcing information, rather than the operator of interest $\boldsymbol{L}$.
The solution to this linear system $\boldsymbol{L}'$ has a growth rate of $\lambda_L-d$ and a frequency of $\pm (\omega_L-cd)$. 
Thus, $\lambda_L-d=0$ is required so that the system maintains a constant amplitude, and the requirement yields the frequency of the LCO $\omega_{\rm NL}=\omega_L-c\lambda_L=1.9$. 
In Fig.~\ref{fig:eval_gain_SL3rd}(\textit{a}), the erroneously estimated eigenvalues $(\lambda_r,\lambda_i)\approx(0.0,\pm 1.9)$ in case of $t\in[0,1500]$ correspond to growth rate and frequency of system $\boldsymbol{L}'$.
This result demonstrates the necessity of the nonlinear forcing snapshots for obtaining $\boldsymbol{L}$.

With regard to the second point (2), let us analyze the proposed FI resolvent operator $\boldsymbol{G}$.
In figure \ref{fig:eval_gain_SL3rd}(\textit{b}), the leading singular values $\sigma_1$ of the resolvent operator $\boldsymbol{R}$ obtained analytically and the proposed data-driven estimation of the FI resolvent operator $\boldsymbol{G}$ are compared.
For computing $\boldsymbol{G}$, snapshots over $t\in[160,1500]$ in the LCO regime are used. 
Since the rank of this system is at most two, as anticipated from the governing equations for two variables \eqref{eq:SL3rd_mat} (though it is actually rank-1 since this model expresses a single oscillator expressed by a single complex variable), the snapshot data are divided into two segments.
For this case, no explicit windowing function for the FFT is employed since the Stuart-Landau oscillator is perfectly periodic.
As a result, as shown in Fig.~\ref{fig:eval_gain_SL3rd}(\textit{b}), the peak of the leading gain of $\boldsymbol{G}$ appears at $\omega=\omega_{\rm NL}$ as expected, while that of the resolvent operator $\boldsymbol{R}$ exists at the frequency of the exponentially growing region $\omega=\omega_L$.
The value of the leading gain at $\omega=\omega_{\rm NL}$ matches the square root of the leading eigenvalue ${\lambda_1^{1/2}(\boldsymbol{S}_{\rm qq}})$ as expected by \eqref{eq:IO_coef}.

\begin{figure}
    \centering
    \includegraphics[width=0.8\linewidth]{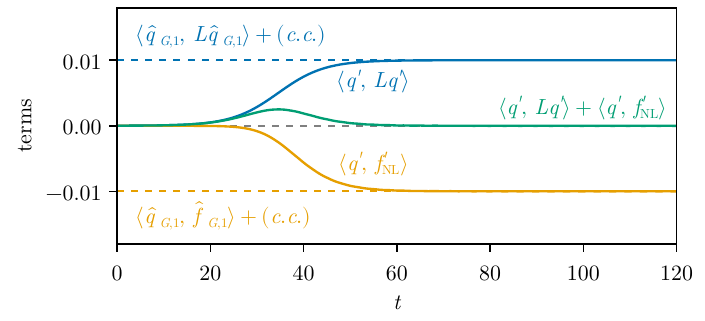}
    \caption{Time evolutions of $\langle q', \boldsymbol{L}q'\rangle$ (solid blue line), $\langle q', f'_{\rm{NL}}\rangle$ (solid orange line), and their sum (solid green line), for the Stuart-Landau system. 
    The dashed blue/orange line is the time-averaged energy supply/removal due to the linear/nonlinear mechanisms during the LCO, evaluated using the leading FI response $\hat{q}_{G,1}$ and forcing modes $\hat{f}_{G,1}$.
    }
    \label{fig:SL3rd_Ebalance}
\end{figure}

Finally, to assess the energy balance between the linear and nonlinear amplification/attenuation mechanisms for the third item (3), the time histories of the terms in \eqref{eq:timeave_evolution_of_Eq'} (before taking time-averaging) are shown in Fig.~\ref{fig:SL3rd_Ebalance}. 
For $t\lesssim60$, the amplification due to the linear term (positive $\langle q',\boldsymbol{L}q'\rangle$ shown in blue line) exceeds the attenuation due to the nonlinear term (negative $\langle q',f'_{\rm{NL}}\rangle$ shown in orange line), resulting in the growth of the amplitude (see also figure \ref{fig:SL3rd_timehist}).
After the system settles into the LCO ($t\gtrsim60$), these two terms are balanced, as indicated by a zero sum (green line). Thus, the amplitude $d$ is kept constant.
Apparently, this plot indicates that the time-averaged values $\overline{\langle q',\boldsymbol{L}q'\rangle}$  and $\overline{\langle q',f'_{\rm{NL}}\rangle}$ during the LCO also balance with each other during the LCO, as indicated by \eqref{eq:timeave_evolution_of_Eq'}.
In addition, the balance between $\langle\hat{q}_{G},\boldsymbol{L}\hat{q}_{G}\rangle+\mathrm{c.c.}$ and $\langle\hat{q}_{G,1},\hat{f}_{G,1}\rangle+\mathrm{c.c.}$ in Fig.~\ref{fig:SL3rd_Ebalance} demonstrates that the energy amplification/attenuation balance for a rank-1 approximated FI resolvent operator in \eqref{eq:energy_balance_lowrank} holds.

As expected, the proposed FI resolvent analysis works for the Stuart-Landau oscillator as an ideal low-dimensional model of oscillatory flow, serving as a foundation for subsequent analyses of actual self-sustained nonlinear flows.

\subsection{Two-dimensional periodic cylinder flow}\label{sec:res_2Dcylinder}
\subsubsection{Numerical simulation set-up}

The proposed framework is then applied to a two-dimensional (2D) cylinder flow at a Mach number $M_\infty=u_\infty/a_\infty=0.2$ and a diameter-based Reynolds number $Re_D=\rho_\infty u_\infty D/\mu_\infty=150$, where $a_\infty$, $\rho_\infty$, $u_\infty$, and $\mu_\infty$, are the speed of sound, density, streamwise velocity, and the dynamic viscosity at freestream.
At the Reynolds number, K{\'a}rman vortex shedding occurs in the cylinder wake.
To obtain the snapshot data, we numerically solve the compressible Navier--Stokes equations in the conservative form.
The computational domain is discretized on a two-dimensional body-fitted O-type grid. The domain extends to $r/D\le100$, where $r$ is the radial distance from the center of the cylinder.
Both radial and circumferential directions have 400 uniformly distributed grid points.
The sixth-order kinetic energy and entropy preserving (KEEP) scheme \citep{Kuya2021-gy} is employed for the convective fluxes, while the sixth-order central differencing is adopted for the viscous fluxes.
The family of the KEEP schemes \citep{Kuya2018-gp,Tamaki2022-hd,Kawai2025-em} achieves non-dissipative while robust computation of compressible flows, and has demonstrated its capability for high-fidelity simulations of a range of self-sustaining nonlinear flows (e.g., \citealp{maejima2025unsupervised,Boldini2025-zu}).
For the time integral, the three-stage total-value-diminishing Runge-Kutta method \citep{Gottlieb1998-ax} is employed, with a Courant number of 0.65.
The no-slip adiabatic condition is imposed at the wall, while the freestream condition is applied for the outer boundary.

\subsubsection{Input-output analysis set-up}

From the numerical simulation, $M=801$ snapshots are obtained with a sampling interval of $\Delta t/(D/u_\infty)\approx 0.27$ after the solution reaches a statistically stationary limit cycle.
The obtained snapshot matrix $\boldsymbol{Q}$ contains 40 cycles of the K\'{a}rman vortex shedding at a Strouhal number of $St=fD/u_\infty\approx0.182$.
Hereafter, let us denote the Strouhal number of the K\'{a}rman vortex shedding by $St_K=0.182$.
Each vortex-shedding cycle contains exactly 20 snapshots.
The nonlinear forcing snapshot matrix $\boldsymbol{F}$ is then computed as a post-processing.
In computing the nonlinear forcing terms, a second-order central difference scheme is used for the spatial derivatives.
On the other hand, for the time derivatives $ \dot{\boldsymbol{Q}}$, the sixth-order central differencing is applied.
According to the modified wave number, the range of $0\le St\le 3.3St_K\sim 4 St_K$ is accurate with the sixth-order central differencing because 5 $\sim$ 6 snapshots per wave are required for resolving a wave with negligible dispersion errors.
In this case, we discuss the results for the first three harmonics, $St=St_K, 2St_K,$ and $3St_K$, which lie within the reliable wavenumber range and contain more than $99\%$ of the fluctuation energy \citep{Jin2021-ge}.

As in \eqref{eq:ROM_L}, the linear operator is estimated using $N_{\rm{POD}}$ POD modes. 
We choose $N_{\rm{POD}}=250$ so that more than $99.9\%$ energy of the original flow field can be recovered. 
On the other hand, to obtain the SPOD modes for the input and output bases, the snapshot datasets are divided into 3 sub-blocks with $50\%$ overlap, yielding $N_p=3$ SPOD modes at each frequency.

To validate the proposed fully data-driven approach, the linear operator $\boldsymbol{L}$ around the mean flow is also computed directly (operator-based approach).
The eigenvalue with the greatest real component of $\boldsymbol{L}$ is obtained using the implicitly restarted Arnold method combined with the shift-inverting method implemented in the Arpack library \citep{lehoucq1998arpack}.
The present eigenvalue is confirmed to be robust to the domain sizes from $r/D\le12.5$ to $r/D\le 32.5$, and the tolerance values for terminating iteration from $10^{-7}$ to $10^{-12}$.
The linear operator $\boldsymbol{L}$ obtained from direct computation is also used to construct $\boldsymbol{G}_{\rm{SDD}}$.

\subsubsection{Results of input-output analysis}

\begin{figure}
    \centering
    \includegraphics[width=0.9\linewidth]{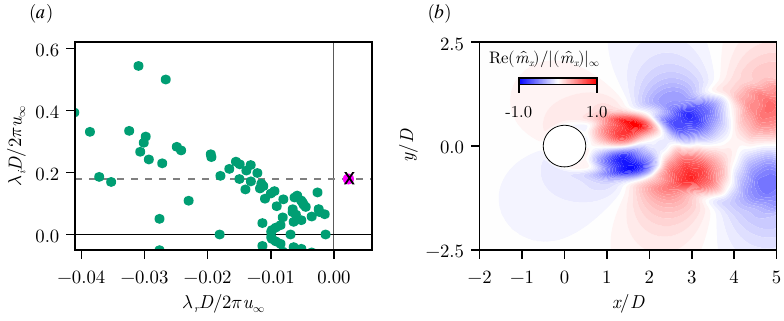}
    \caption{Data-driven estimate of linear time-invariant operator $\boldsymbol{L}_{\rm{data}}$ of two-dimensional cylinder flow.
    (a) Eigenvalues $\lambda=\lambda_r+\mathrm{i}\lambda_i$ of the proposed $\boldsymbol{L}_{\rm{data}}$ (green and magenta circles) and the most unstable eigenvalue of $\boldsymbol{L}$ computed in operator-based approach (black cross), (b) real part of the streamwise momentum component of the eigenvector $\hat{m}_x$ of $\boldsymbol{L}_{\rm{data}}$ corresponding to the most unstable eigenvalue indicated by the magenta circle. 
    }
    \label{fig:eigsL_cylinder2D}
\end{figure}

\begin{figure}
    \centering
    \includegraphics{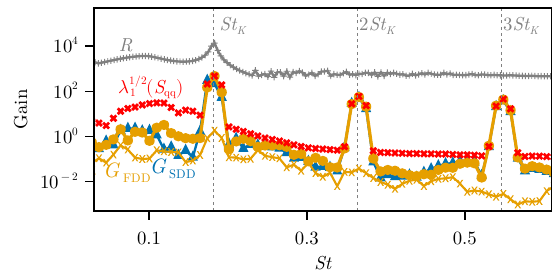}
    \caption{Leading gains (singular values $\sigma_1$) for 2D cylinder case, obtained from the fully data-driven $\boldsymbol{G}_{\rm{FDD}}$ (orange circle), semi-data-driven $\boldsymbol{G}_{\rm{SDD}}$ (blue triangle), and the ordinary resolvent operator $\boldsymbol{R}$ (gray plus), along with the secondary leading gains of $\sigma_2(\boldsymbol{G}_{\rm{FDD}})$(orange cross). 
    The square roots of the leading eigenvalues SPOD gains $\boldsymbol{S}_{\rm{qq}}$(red cross) are also shown.
    }
    \label{fig:cylinder_gain}
\end{figure}

Let us first look at the eigenvalues of the linear operator estimated from data $\boldsymbol{L}_{\rm{data}}$.
The obtained eigenvalue with the greatest real part (magenta circle) agrees well with the one obtained by the operator-based approach (black cross), as shown in Fig.~\ref{fig:eigsL_cylinder2D}(\textit{a}).
The corresponding eigenvector in Fig.~\ref{fig:eigsL_cylinder2D}(\textit{b}) indicates the physical mode associated with the K\'{a}rman vortex shedding.
The results demonstrate that the linear operator $\boldsymbol{L}$ is well estimated by the proposed method.

The leading singular values $\sigma_1$ of the FI resolvent operators $\boldsymbol{G}_{\rm{FDD}}$ (orange) and $\boldsymbol{G}_{\rm{SDD}}$ (blue) are shown in Fig.~\ref{fig:cylinder_gain}, along with the conventional resolvent gain $\sigma_1(\boldsymbol{R})$.
The leading gains of both $\boldsymbol{G}_{\rm{FDD}}$ and $\boldsymbol{G}_{\rm{SDD}}$  possess the peaks not only at $St_K$ but also its harmonics $2St_K, 3St_K$, indicating that the present FI resolvent operator captures the nonlinear amplification mechanisms which the conventional resolvent $\boldsymbol{R}$ does not predict.
This result indicates that the output energy at each frequency near the gain peaks is successfully recovered from both the fully data-driven estimates $\widetilde{\boldsymbol{G}}_{\rm{FDD}}$ and the semi-data-driven estimates $\widetilde{\boldsymbol{G}}_{\rm{SDD}}$.
The correspondence of the leading gains between $\widetilde{\boldsymbol{G}}_{\rm{FDD}}$ and $\widetilde{\boldsymbol{G}}_{\rm{SDD}}$ also supports the validity of the estimated linear operator $\boldsymbol{L}_{\rm{data}}$ used in the fully data-driven framework.
Moreover, near the gain peaks at the frequencies of interest ($St_K, 2St_K$, and $3St_K$), the leading gains $\sigma_1$ show a good agreement with the square root of the leading SPOD gains of output, i.e., $\sigma_1(\boldsymbol{G})\approx\lambda_1(\boldsymbol{S}_{qq})^{1/2}$.
This is consistent with \eqref{recover_output_amp} for a rank-1 FI resolvent operator, which holds at the harmonics where the leading and secondary gains are separated by about two orders of magnitude.
Slight differences between the FI resolvent and SPOD gains at frequencies other than the harmonics of $St_K$ may be observed, but they are not the main concern of this analysis.

\begin{figure}
    \centering
    \includegraphics[width=0.99\linewidth]{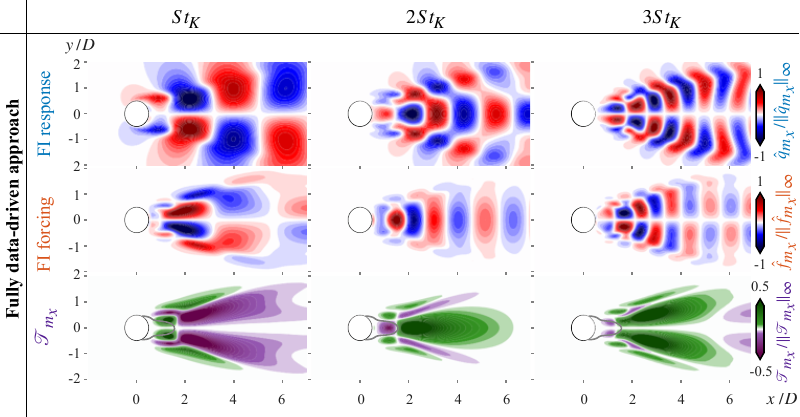}
    \caption{Spatial distributions of the leading FI response and forcing modes in 2D cylinder wake extracted by the fully data-driven approach using $\boldsymbol{G}_{\rm{FDD}}$. 
    Real parts of streamwise momentum components are shown for the FI response $\hat{q}_{m_x}/\|\hat{q}_{m_x}\|_\infty$ (top) and forcing $\hat{f}_{m_x}/\|\hat{f}_{m_x}\|_\infty$ (middle). 
    The nonlinear energy transfer map for the streamwise momentum component $\mathcal{T}_{m_x}:=\mathrm{Re}[\boldsymbol{Z}\hat{q}_{m_x}^*\odot \boldsymbol{Z}\hat{f}_{m_x}]$, scaled by its infinity norm $\|\mathcal{T}_{m_x}\|_\infty$, is shown at the bottom.}
    \label{fig:mode_data_cylinder2D}
\end{figure}

\begin{figure}
    \centering
    \includegraphics{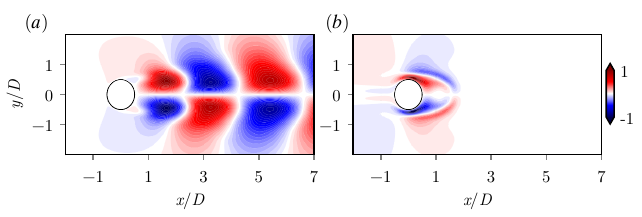}
    \caption{Conventional resolvent modes obtained using $\boldsymbol{R}$: streamwise component of (a) response mode, (b) forcing mode. The real part of the momentum component scaled by its infinity norm is shown.}
    \label{fig:ordinary_resolvent}
\end{figure}

The streamwise momentum component of the FI response and forcing modes obtained by the fully data-driven approach are shown in Fig.~\ref{fig:mode_data_cylinder2D}.  
Hereafter, we use $\hat{q}_{m_x}$ and $\hat{f}_{m_x}$ to denote the streamwise momentum component of the leading FI response and forcing modes $\hat{q}_{G,1}$ and $\hat{f}_{G,1}$, respectively. 
We first emphasize that these FI response and forcing modes are grounded in the actual flow field, thanks to the use of the basis vectors of the input and output subspaces, even at harmonics $2St_K$ and $3St_K$, where the nonlinear amplification mechanism is crucial. 
For comparison, we also visualize the response and forcing modes obtained by the conventional resolvent analysis in Fig.~\ref{fig:ordinary_resolvent}.
Looking at the FI response mode at $St_K$ in Fig.~\ref{fig:mode_data_cylinder2D}, the spatial structure of the FI response mode resembles the conventional response mode shown in Fig.~\ref{fig:ordinary_resolvent}(\textit{a}).
In contrast, the FI forcing mode structure differs greatly from the conventional resolvent forcing mode \ref{fig:ordinary_resolvent}(\textit{b}): the FI forcing mode is confined to the cylinder wake, whereas the conventional forcing mode appears not only near the shear layer but also upstream of the cylinder.
The FI forcing mode that sustains oscillation of $\hat{q}$ should be confined to the place where the output $q'$ appears because the ${f'}_{\rm{NL}}$ is a function of $q'$.
In this sense, the proposed FI forcing mode reveals the nonlinear forcing component that plays a crucial role in the self-sustaining mechanism, which is not captured well by the conventional forcing mode.
In a similar manner, the FI forcing modes at the harmonics $2St_K$(middle) and $3St_K$(right) also appear in the wake where the FI response modes exist.

To explore the role of the FI forcing mode at the fundamental frequency $St_K$ in the self-sustaining mechanism, we visualize the streamwise momentum component of the nonlinear energy transfer map $\mathcal{T}_{m_x}:=\mathrm{Re}[\boldsymbol{Z}\hat{q}_{m_x}^*\odot \boldsymbol{Z}\hat{f}_{m_x}]$ from \eqref{eq:NLenergyTransferMap} in Fig.~\ref{fig:mode_data_cylinder2D}(bottom).
As discussed earlier, in the self-sustained flows, the time-averaged energy balance between the amplification/attenuation via the linear and nonlinear mechanisms  \eqref{eq:timeave_evolution_of_Eq'} holds, and we can consider the nonlinear energy map $\mathcal{T}$ especially at the frequency where the rank-1 approximation of the FI resolvent operator is valid.
In figure \ref{fig:mode_data_cylinder2D}, the nonlinear energy transfer map $\mathcal{T}_{mx}$ identifies the local spatial domain where the energy amplification/attenuation mechanisms due to the nonlinear forcing term are prevalent, as indicated by the green region for amplification and by the purple region for attenuation.
The purple region of $\mathcal{T}$ at $St_K$ in Fig.~\ref{fig:mode_data_cylinder2D} indicates that the streamwise momentum component of the FI forcing mode attenuates the fluctuation energy at the $St_K$ scale in the cylinder wake, except for the green region, including near the recirculation region (gray line) right behind the cylinder and the interference region with the free stream. 
In contrast, at the harmonics ($2St_K,3St_K$), the FI forcing mode attenuates the energy near the recirculation region while supplying it to most other regions in the wake. 
These maps of $\mathcal{T}_{m_x}$ suggest that the fluctuation energy at $St_K$ is generated near the recirculation region (green region at $St_K$) and then transported downstream by convection, followed by the interscale energy transfer by the nonlinear forcing $f'_{\rm}$ that supplies the energy to the harmonics in the downstream of the cylinder wake ($2\lesssim x/D\lesssim 6$) (purple region at $St_K$ whereas green regions at $2St_K$ and $3St_K$).
Focusing on the positive energy transfer by the nonlinear forcing near the recirculation region at $St_K$, this result indicates that the nonlinear forcing associated with shear-layer roll-up dynamics at $St_K$ continues to supply energy to the streamwise momentum oscillation, through the mixing of the streamwise moment in the transverse direction $y$ as represented by the fluctuation component of $-\overline{\rho}^{-1} \partial (m_x'm_y')/\partial y$ in the nonlinear forcing $f'_{\rm{NL}}$ \eqref{eq:fnl_implemtented}.
As demonstrated here, the spatial distributions of the nonlinear energy transfer $\mathcal{T}$ locate the region where the FI forcing mode supplies/removes the fluctuation energy, which aids in the physical interpretation of the FI forcing mode.
We note that a brief qualitative assessment of the nonlinear energy transfer map, including other variables, is provided in Appendix.\ref{app:energymap}.
In this assessment, we show that the FI response and forcing modes capture the nonlinear energy transfer component that is not captured by conventional resolvent analysis \citep{Jin2021-ge}.

\begin{figure}
    \centering
    \includegraphics[width=0.8\linewidth]{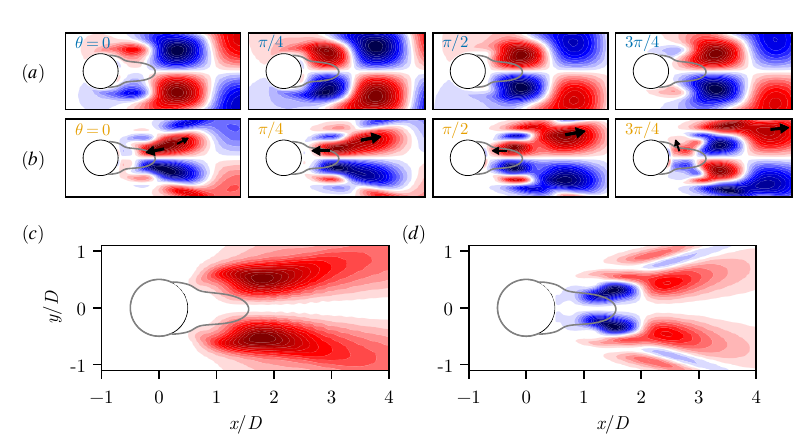}
    \caption{Phase evolutions of the FI (\textit{a}) response and (\textit{b}) forcing modes at $St_K$, along with the (normalized) streamwise component of phase velocity of FI (\textit{c}) response and (\textit{d}) forcing modes.
    In (\textit{b}), the black arrows indicate the direction of the movement of the FI forcing mode structure.
    Spatial distributions of $-\nabla \angle \hat{q}_{G,m_x}/\|\nabla \angle \hat{q}_{G,m_x}\|$ and $-\nabla \angle \hat{f}_{G,m_x}/\|\nabla\angle \hat{f}_{G,m_x}\|$ are shown in (\textit{c}) and (\textit{d}), respectively. The gray isoline of $\overline{u}=0$ indicates the recirculation domain. }
    \label{fig:forcing_flow_direction}
\end{figure}

It is also interesting to note that, from the time (phase) evolution of the FI response and forcing modes in Fig.~\ref{fig:forcing_flow_direction}(\textit{a},\textit{b}), we observe that only the FI forcing mode at $St_K$ near the recirculation region propagates upstream (see black arrows).
To quantitatively show that, we visualize the streamwise component of the normalized phase velocity in Fig.~\ref{fig:forcing_flow_direction}(\textit{c},\textit{d}), as in \citet{Poplingher2019-kp}.
These figures indicate that the FI response mode convects downstream, whereas the FI forcing mode in the recirculation region travels upstream.
In other words, the FI forcing mode transmits disturbances upstream and forms spatially localized feedback, which is a key feature of self-sustained flows according to \citet{Schmid2016-xw}, who noted that the strength of feedback can classify a self-sustained flow into oscillatory flow or amplifier flow. 
Since this spatially localized feedback never stops once it occurs, we argue that the nonlinear dynamics near the recirculation region are a core energy source of the self-sustained vortex shedding in the cylinder wake.

\begin{figure}
    \centering
    \includegraphics{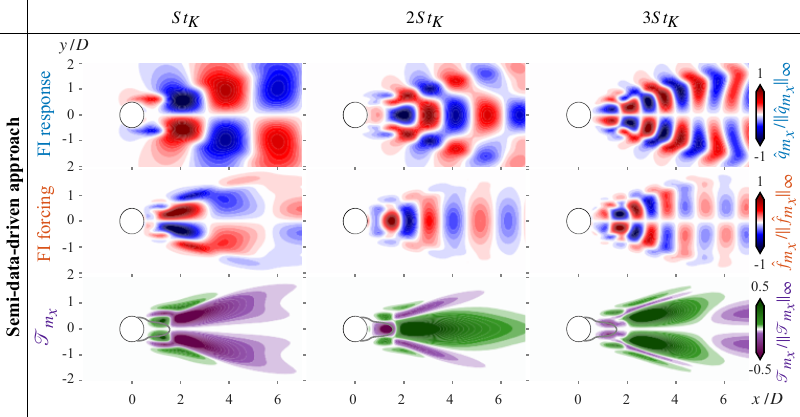}
    \caption{Same as figure \ref{fig:mode_data_cylinder2D} but for semi-data-driven approach using $\boldsymbol{G}_{\rm{SDD}}$.}
    \label{fig:mode_optl_cylinder2D}
\end{figure}

Finally, we also show that the result obtained using the operator in a semi-data-driven approach $\boldsymbol{G}_{\rm{SDD}}$ in Fig.~\ref{fig:mode_optl_cylinder2D}.
The obtained optimal response and forcing distributions are consistent with those in the fully data-driven case shown above.
The consistent results support the validity of the results obtained by the fully data-driven approach, as well as the semi-data-driven approach.

\subsection{Laminar-turbulent transition in boundary layer}\label{sec:3.3}

\subsubsection{Case description and numerical simulation set-up}
\begin{figure}
    \centering
    \includegraphics[width=\linewidth]{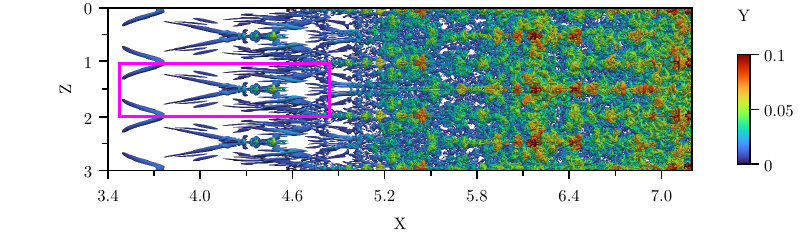}
    \caption{Topview of the instantaneous iso-surfaces of Q criterion $Q/(a_\infty/x_0)^2=25$ colored by the height from the wall $Y:=y/x_0$ of the DNS data \citep{Iwatani2025-ty}, where $x_0$ is the reference length.
    The magenta rectangle domain is used for the FI resolvent analysis.
    Here, $X:=(x-x_d)/x_0$ and $Z:=z/\Lambda_{z}$, where $x_d$ is the disturbance slot location and $\lambda_z$ is the wavelength of the subharmonic initial disturbance.
    }
    \label{fig:DNSoverall}
\end{figure}

Finally, we perform the fully data-driven FI resolvent analysis of the laminar-turbulent transition in the boundary layer, with particular focus on the late stage where the three-dimensional (3D) vortex structures evolve nonlinearly.
The DNS data of the spatially developing boundary layer over a nominally zero-pressure-gradient, pseudo-adiabatic (isothermal) flat plate are prepared \citep{Iwatani2025-ty},  as shown in Fig.~\ref{fig:DNSoverall} (the magenta rectangle domain is used for the FI resolvent analysis as detailed later).
The DNS solves the entire process of the spatially developing transitional boundary layer, covering from the linear growth of the instability to the equilibrium turbulent boundary layer.
The freestream Mach number is $0.8$, and the reference Reynolds number is $Re_{x_0}\equiv \rho_\infty u_\infty x_0/\mu_\infty=10^5$, where $x_0$ is the reference streamwise location such that $x/x_0=Re_x/10^5$.
The DNS employs the 6th-order compact differencing scheme \citep{Lele1992-hz} for the spatial discretization and the four-stage fourth-order Runge-Kutta method for the time discretization with a Courant number of 0.6.

In the DNS, the initial blowing-suction type disturbances are introduced from the disturbance slot $x_d/x_0=3.4$ on the wall to trigger instabilities following the prior DNS studies of controlled transition \citep{Sayadi2013-ub,Boldini2025-zu}.
This study considers the H-type transition, triggered by a planar (2D) Mack's first mode \citep{Mack1975-oc} as the primary instability mode and a 3D subharmonic oblique mode \citep{Herbert1988-qu} as the secondary instability mode.
The Mack's first mode has frequency $\omega_{2D}$, whereas the subharmonic oblique mode has half the frequency of $\omega_{2D}$, i.e., $\omega_{2D}/2$ and spanwise wavelength $\lambda_z$.
With this DNS setup, Mack's first mode exhibits linear growth, and its secondary instability arises from disturbance of the subharmonic oblique mode.
As a result, the 3D vortices, such as $\Lambda$-vortex, hairpin vortex, and ring-vortex, emerge and form a staggered arrangement, appearing alternately at the spanwise locations of $Z:=z/\lambda_z\approx N$ or $Z\approx 1/2 + N$ ($N=0,1,2$).
The staggered arrangement of the 3D vortices is characteristic of the late stage of the transition triggered by the subharmonic secondary instability  \citep{Herbert1988-qu}, typically observed in the H-type transition.
The focus here is on the nonlinear mechanism at this late stage of the H-type transition, rather than the linear amplification mechanism of Mack's first mode and the subharmonic secondary instability mode.
The reader is referred to \citet{Iwatani2025-ty} for details of the DNS.

\begin{figure}
    \centering
    \includegraphics[width=0.95\linewidth]{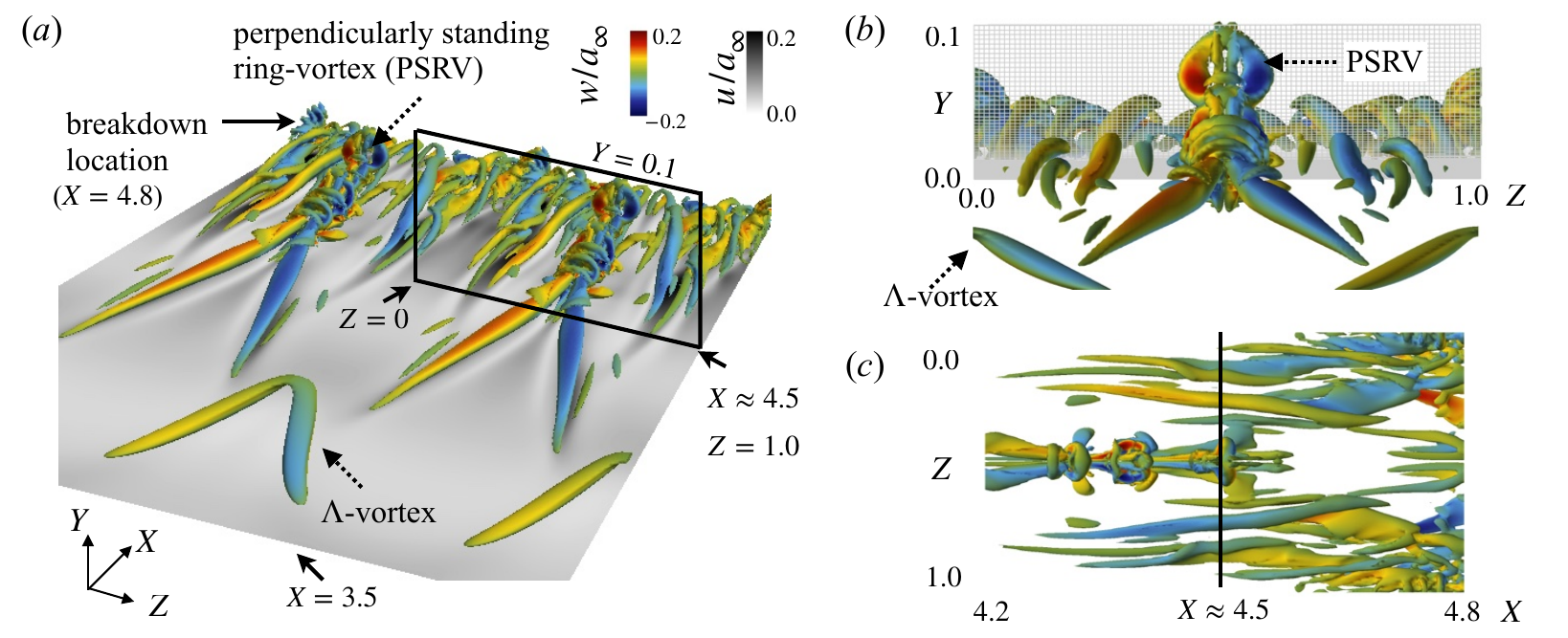}
    \caption{Late stage of H-type transition, visualized by iso-surfaces of Q criterion ($Q/(a_\infty/x_0)^2=10$) colored by spanwise velocity $w$: (\textit{a}) oblique view of the whole analysis domain (copied once in the spanwise direction $Z$), and (\textit{b}) front and (\textit{c}) top views near the perpendicularly standing ring-vortex (PSRV) denoted in black rectangle in (\textit{a}).  
    }
    \label{fig:transition-field}
\end{figure}

\subsubsection{FI resolvent analysis set-up}
For the fully data-driven FI resolvent analysis, spatially and temporally sub-sampled snapshot data are collected.
The DNS volume data of $M=1536$ snapshots are collected with a sampling time interval of $\Delta t/(a_\infty/x_0)=0.018$, corresponding to a sampling (angular) frequency of almost $16\omega_{2D}$.
This time-series dataset holds precisely 22 cycles of Mack's first mode (11 cycles of the subharmonic oblique mode).
Since the eighth-order central differencing method, requiring four snapshots for resolving a wave, is used for the computation of the time-derivative term, the reliable range of frequency is $|\omega/\omega_{2D}|\le 4$.

As shown in Fig.~\ref{fig:transition-field}(\textit{a}), the analysis domain is defined as follows.
The streamwise range $X:=(x-x_d)/x_0$ ($X=0$ is the disturbance slot location) extends from the onset of the 3D $\Lambda$-vortices ($X=3.5$) to the boundary-layer breakdown $X=4.8$.
Here, we define the breakdown location as the streamwise station at which the broadband spanwise spectra of flow quantities first appear.
The wall-normal domain $Y:=y/x_0$ ranges from the wall to a point sufficiently far from the wall to involve all vortex dynamics.
The spanwise domain $Z$ is set to $0\le Z\le 1$, which is sufficiently wide to capture 3D vortices.
Also, the original DNS data is spatially sub-sampled with every second grid point in the streamwise direction and every third point in both the wall-normal and spanwise directions, resulting in $408 \times 105 \times100 \approx 4\times10^6$ grid points.
Nevertheless, the resultant grid spacings are fine enough to resolve the flow structures observed in the range of $|\omega/\omega_{2D}|\le 4$, using the employed six-order compact differencing scheme for computing the nonlinear forcing term.

To obtain the SPOD modes for the input and output basis vectors, the time series data are subdivided into 10 blocks with 50\% overlap, each block holding 4 cycles of Mack's first mode $\omega_{2D}$.
The number of POD modes for obtaining $\boldsymbol{L}_{\rm{data}}$ is determined in the next section after checking the convergence.

In the analysis domain, the ring vortices develop downstream and eventually stand perpendicular to the wall at $X\approx 4.5$ (see closed-up view in Fig.~\ref{fig:transition-field}(\textit{b},\textit{c})).
This perpendicularly standing ring vortex (PSRV) is reported to be a key vortex structure in boundary layer breakdown \citep{Lu2012-bh} as it generates a high-shear region through intense momentum exchange induced by strong ejection and sweep motions under it.
Because of its physical importance, we focus on the PSRV in the FI resolvent analysis later when discussing the mode distribution of the wall-normal--spanwise ($Y$--$Z$) cross-section.

\subsubsection{Results of fully data-driven FI resolvent analysis}
\begin{figure}
    \centering
    \includegraphics[width=0.8\linewidth]{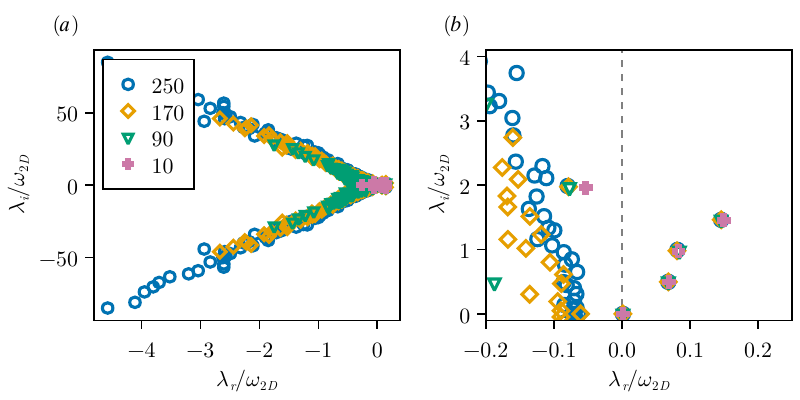}
    \caption{Eigenvalues $\lambda_r+\mathrm{i}\lambda_i$ of the estimated linear operator $\boldsymbol{L}_{\rm{data}}$ obtained using $N_{\rm{POD}}=10,~90,~170,250$:  (\textit{a}) overall view; (\textit{b}) enlarged view near unstable eigenvalues.
    }
    \label{fig:eval_transition}
\end{figure}

\begin{figure}
    \centering
    \includegraphics[width=0.95\linewidth]{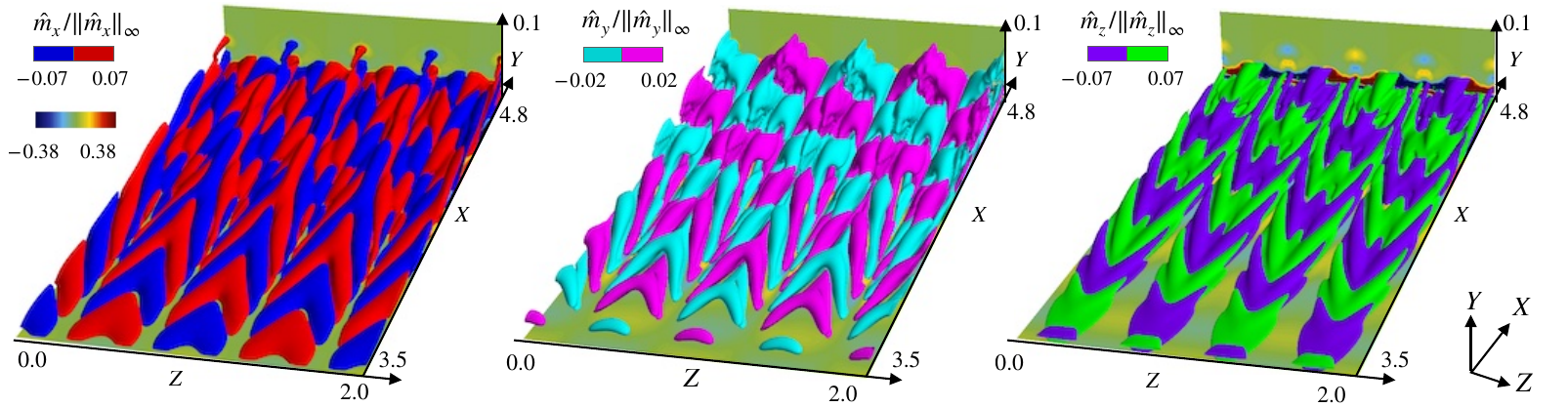}
    \caption{
    Isosurfaces of momentum components $\hat{m}$ for the most unstable eigenmodes of the H-type transitional boundary layer ($\lambda_i/\omega_{2D}\approx 3/2$), computed from the data-driven linear operator $\boldsymbol{L}_{\rm{data}}$ with $N_{\mathrm {pod}}=90$: (left) streamwise $\hat{m}_x$, (middle) wall-normal $\hat{m}_y$, (right) spanwise $\hat{m}_z$ momentum components.
    For visualization, the iso-surfaces are copied once in the spanwise direction.
    }
    \label{fig:evec_transition}
\end{figure}

The eigenvalues $\lambda_r+\mathrm{i}\lambda_i$ of the data-driven estimated linear operator $\boldsymbol{L}_{\rm{data}}$ obtained using $N_{\rm{POD}}=10,90,180,250$ POD modes are shown in Fig.~\ref{fig:eval_transition}.
All of those cases satisfies $\sum_{j}^{N_{\rm{POD}}}\sigma_j^2(X)/\sum_{j}^M\sigma_j^2(X)\ge0.999$.
In this figure, the convergence of the unstable eigenvalues with respect to the number of POD modes $N_{\rm{POD}}$ is confirmed for $N_{\rm{POD}}\ge10$.
Hereafter, the $N_{\rm{POD}}=90$ case is discussed, at which a stable eigenvalue at $(\lambda_r/\omega_{2D}, \lambda_i/\omega_{2D})\approx(-0.08, 2)$ is also converged.
The eigenvalues with positive real part $\lambda_r$ of $\boldsymbol{L}_{\rm{data}}$ indicate that the linear energy amplification mechanism is predominant at frequency $0<|\lambda_i/\omega_{2D}|\lesssim3/2$.
On the other hand, at frequencies $|\lambda_i/\omega_{2D}|\ge 1.5$, the real part of the eigenvalues $\lambda_r$  is negative, indicating that linear amplification mechanism inactive, and that is, the nonlinear amplification mechanism is predominant.
Note that the linear amplification mechanisms at the $\lambda_i/\omega_{2D}\approx1,1/2$ are different from the growth mechanism of Mack's first mode and subharmonic secondary instability mode, respectively.
As shown in Fig.~\ref{fig:evec_transition}, the momentum components of the eigenvector that corresponds to the eigenvalue with the greatest real part at $\lambda_i/\omega_{2D}\approx3/2$ also exhibit physical staggered structures. 
The pairs of spanwise-aligned opposite-signed structures (red-blue, magenta-cyan, purple-green) appear and develop downstream, changing their signs alternately.
These modal structures are reminiscent of the staggered alignment of the 3D vortices in the late stage of the H-type transition \citep{Herbert1988-qu}.
As such, the data-driven estimation of the linear operator captures the flow physics well.


\begin{figure}
    \centering
    \includegraphics[width=0.8\linewidth]{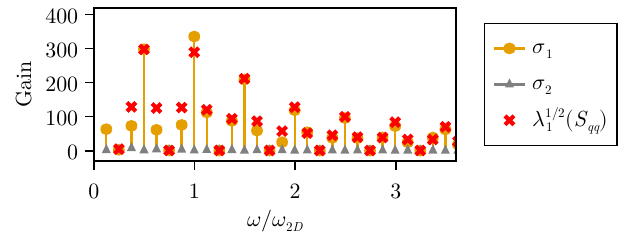}
    \caption{Gains, $\sigma_{G,1}$ and $\sigma_{G,2}$, of the FI resolvent operator $\boldsymbol{G}_{\rm{FDD}}$ for the transition boundary layer case, compared with the square root of the leading SPOD gains $\lambda_1^{1/2}(\boldsymbol{S}_{\mathrm{qq}})$.}
    \label{fig:transition-gain}
\end{figure}

Let us move on to the FI resolvent analysis using the estimated linear operator.
First, looking at the leading gains shown in Fig.~\ref{fig:transition-gain}, the leading gains at harmonics of $(\omega_{2D}/2)N~(N=1,2,\dots)$ collapse well onto the corresponding leading SPOD gains $\lambda^{1/2}_{1}(\boldsymbol{S}_{\rm{qq}})$ (the output amplitudes), ensuring the validity of the fully data-driven FI-resolvent analysis of this case.
The leading gains exhibit peaks not only at the frequencies $0<|\omega/\omega_{2D}|\lesssim1.5$ where the linear amplification mechanism is active (as seen in the eigenvalues of $\boldsymbol{L}_{\rm{data}}$), but also at the frequencies $|\omega/\omega_{2D}|\ge 1.5$ where the nonlinear linear amplification mechanism is predominant.
The secondary leading gains (gray triangles) show a large gap between the first leading ones, especially at the frequencies $(\omega_{2D}/2)N$ where the high gain (amplitude) is obtained.
This result indicates the rank-1 property of the FI resolvent operator at these dominant frequencies in the late stage of the H-type transition. 
Although the gain at $\omega/\omega_{2D}=1$ is slightly overestimated, its order is still the same as the SPOD gain.

\begin{figure}
    \centering
    \includegraphics[width=0.95\linewidth]{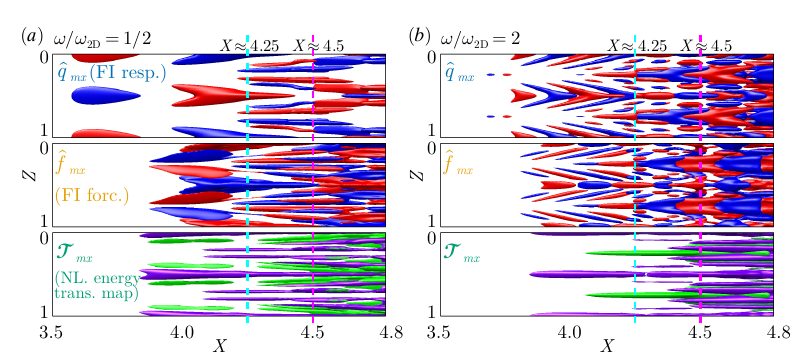}
    \caption{Results of FI resolvent analysis of transitional boundary layer for streamwise momentum component at (\textit{a}) $\omega/\omega_{2D}=1/2$ and (\textit{b}) $2$. 
    Topviews of iso-surfaces of  the FI response mode $\hat{q}_{m_x}/|\hat{q}_{m_x}|_\infty=\pm 0.25$ (top), the FI forcing mode $\hat{f}_{m_x}/\|\hat{f}_{m_x}\|_\infty=\pm 0.25$ (middle), and the nonlinear energy transfer map $\mathcal{T}_{m_x}/\|\mathcal{T}_{m_x}\|_\infty=\pm 0.25$ in (\textit{a}) and $0.1$ in (\textit{b})(bottom). 
    Red and green indicate the positive iso-values, whereas blue and purple indicate the negative iso-values.
    }
    \label{fig:FImode-transition3D}
\end{figure}

Subsequently, we look at the FI response and forcing modes for the streamwise momentum component.
In Fig.~\ref{fig:FImode-transition3D}, we focus on the FI resolvent modes at $\omega/\omega_{2D}=1/2$ and $2$, as representative frequencies for the predominant linear and nonlinear amplification mechanisms, respectively. 
More specifically, at $\omega/\omega_{2D}=1/2$, the linear instability mechanism is predominant and the nonlinear forcing suppresses the energy to maintain the energy balance~\eqref{eq:energy_balance}, whereas the roles of linear and nonlinear energy amplification/attenuation mechanisms switch at $\omega/\omega_{2D}=2$.
Observing Fig.~\ref{fig:FImode-transition3D}, the FI response and forcing modes at these frequencies identify coherent structures that develop downstream into more complex geometries, associated with the downstream development of the transitional boundary layer.
In particular, at both frequencies, the FI response and forcing modes evolve into complex structures near the breakdown region at $4.3\lesssim X\lesssim 4.8$ (downstream of cyan dashed line), which contains the PSRV.
The structures also become more complex in the spanwise direction. 
The FI response and forcing modes emerge simultaneously in the spanwise gap between the staggered 3D vortices (i.e., $Z\approx 1/4, ~3/4$), which reflects the fact that the FI forcing is a nonlinear function of the output $q'$. 

The maps of $\mathcal{T}_{m_x}$ uncover the coherent structure of the nonlinear energy transfer of the streamwise momentum component, latent in the late stage of the H-type transition, as shown by green (positive) and purple (negative) iso-surfaces in Fig.~\ref{fig:FImode-transition3D}. 
The nonlinear energy transfer mechanisms at the two frequencies $\omega/\omega_{2D}=1/2$ and $2$ share similarities and differences, as explained below.
A common feature is that the purple structures of $\mathcal{T}$ of both frequencies are observed at $Z=0,1/2$ and $1$, each aligns with the centerline between the legs of the staggered 3D $\Lambda$-shaped and ring-like vortices (see also figures~\ref{fig:DNSoverall} and \ref{fig:transition-field}. 
This indicates that the FI forcing at these spanwise locations attenuates the fluctuation energy of these 3D vortices, at least on the visible far-wall portion in this top view (we will see the near-wall structure later).
On the other hand, the sign of the energy transfer map is opposite between the two frequencies in the spanwise gap between the staggered 3D vortices $Z\approx 1/4$ and $3/4$ near the breakdown region.
The FI forcing at $\omega/\omega_{2D}=1/2$ in Fig.~\ref{fig:FImode-transition3D}(\textit{a}) attenuates the energy at this region, whereas that at $\omega/\omega_{2D}=2$ in Fig.~\ref{fig:FImode-transition3D}(\textit{b}) supplies the energy to this region.
This result suggests that the nonlinear forcing at $\omega/\omega_{2D}=2$ provides the energy for the formation of the spanwise complex structures in the boundary layer before the breakdown to turbulence.

\begin{figure}
    \centering
    \includegraphics[width=0.95\linewidth]{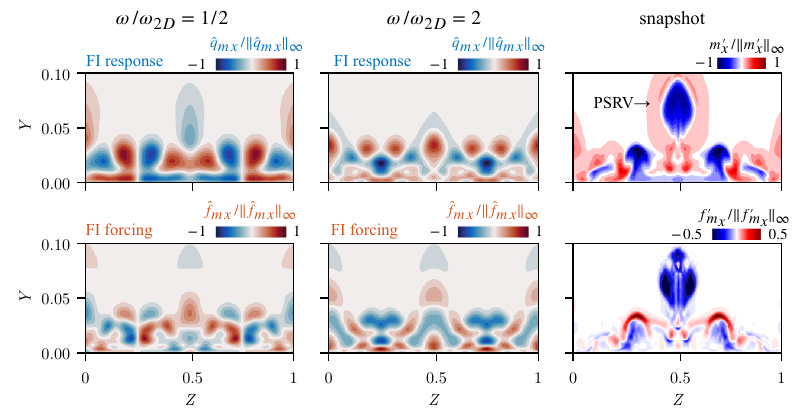}
    \caption{
    Input-output relations at $X\approx 4.5$ where a perpendicularly standing ring-like vortex (PSRV) is observed. Streamwise momentum component of the FI response and forcing modes at $\omega/\omega_{2D}=1/2$ (left) and $2$ (middle). Snapshots of streamwise momentum component of the output $m'_x$ and nonlinear forcing $f'_{m_x}$ (right).
    }
    \label{fig:transition-FImodes-snapshot}
\end{figure}

We further examine the extracted input-output relations at a streamwise location $X\approx 4.5$ (the magenta dashed line in Fig.~\ref{fig:FImode-transition3D}), at which the PSRV is observed in $Y\gtrsim 0.05$ (see Fig.~\ref{fig:transition-field}).
As shown in Fig.~\ref{fig:transition-FImodes-snapshot}, the streamwise momentum component of the FI response and forcing modes at both frequencies at $\omega/\omega_{2D}=1/2$ and $2$ extract the coherent structures under the PSRV ($Y\lesssim 0.05$).
The FI response mode at $\omega/\omega_{2D}=1/2$, which is a dominant component as indicated by a large amplitude $\sigma_1(\boldsymbol{G})$ in Fig.~\ref{fig:transition-gain}, shows similar structures to the instantaneous flow structures of streamwise momentum fluctuation $m_x'$ around the PSRV and below it at $0.3\lesssim Z\lesssim 0.7$.
This supports that the FI response extracts the flow structures present in the actual flow field.
The FI forcing mode also isolates the key component for the self-sustaining mechanism from the region where the nonlinear forcing snapshot $f'_{m_x}$ (as a function of $q'$) is present.

\begin{figure}
    \centering
    \includegraphics[width=0.9\linewidth]{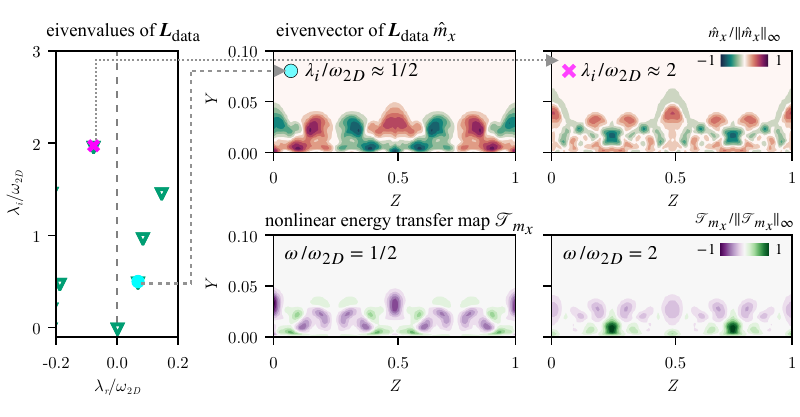}
    \caption{
    Comparison between the eigenmode $m_x$ (global stability mode) of $\boldsymbol{L}_{\mathrm{data}}$ using $N_{\rm{POD}}=90$  and nonlinear energy transfer map $\mathcal{\boldsymbol{T}}_{m_x}$, for streamwise momentum component, at $\omega/\omega_{2D}=1/2$ and $2$. 
    Eigenvalues $\lambda_r+\mathrm{i}\lambda_i$ of $\boldsymbol{L}_{\rm{data}}$ indicate the stability of the corresponding eigenmodes.
    }
    \label{fig:transition-Tmap-Lmode}
\end{figure}

To gain further insights into the physical role of the extracted FI forcing modes, the nonlinear energy transfer maps at $\omega/\omega_{2D}=1/2$ and $2$ are shown in Fig.~\ref{fig:transition-Tmap-Lmode}.
In the same Fig.~\ref{fig:transition-Tmap-Lmode}, the eigenmodes of these frequencies $\lambda_i/\omega_{2D}\approx1/2$ and $2$ are also shown, along with the corresponding eigenvalues, highlighted by the cyan circle and magenta cross, respectively. 
Hereafter, we use $\omega/\omega_{2D}$ instead to denote the frequency of these eigenmodes.
It is evident from the real part of these eigenvalues that the eigenmode at $\omega/\omega_{2D}\approx1/2$ (cyan circle) is linearly unstable, whereas that $\omega/\omega_{2D}\approx1/2$ (magenta cross) is linearly stable.
Comparing the structures of these eigenmodes and the nonlinear energy transfer map at $\omega/\omega_{2D}=1/2$, some local peaks of $\mathcal{T}$ (dark purple regions) appears in the region where the eigenmode attains a large amplitude (dark red/green).
This positional relationship indicates that the FI forcing attenuates the fluctuation energy amplified by the linear instability to prevent unlimited linear energy growth, thereby maintaining oscillation energy at a statistically stationary level.
Conversely, at $\omega/\omega_{2D}\approx2$, the eigenmode is supposed to decay eventually in terms of the linear stability. 
However, as the energy transfer map in Fig.~\ref{fig:transition-Tmap-Lmode} reveals, the FI forcing mode injects the energy into the domain at $Z\approx 1/4$ and $3/4$ (the spanwise gap between the PSRV) and $ Y\lesssim 0.02$ (near the wall), which have been mentioned also in the iso-surface plot in Fig.~\ref{fig:FImode-transition3D}(\textit{b}). 
By supplying the fluctuation energy, the FI forcing allows the linearly stable eigenmode to persist, resulting in the FI response similar to the eigenmode (compare the FI response in Fig.~\ref{fig:transition-FImodes-snapshot} and the eigenmode in Fig.~\ref{fig:transition-Tmap-Lmode} at $\omega/\omega_{2D}=2$).
We also note that, as we observe the green region of $\mathcal{T}_{m_x}$ near the wall under the PSRV at $Z\approx 0.5$ in Fig.~\ref{fig:transition-Tmap-Lmode}, the FI forcing supplies the energy to the region at both $\omega/\omega_{2D}=1/2$ and $2$, which was obscured in the top view of the iso-surface visualization of $\mathcal{T}_{m_x}$ in Fig.~\ref{fig:FImode-transition3D} and is now evident from this cross-section view.
Considering the discussion of the PSRV as a key structure for turbulent breakdown \citep{Lu2012-bh}, the positive nonlinear energy transfer near the wall suggests a possible connection between nonlinear, intense sweep motions under the PSRV and the fluctuation energy amplification associated with the turbulent breakdown.

As demonstrated in the transitional boundary layer, the present FI resonant analysis and interpretation of the FI forcing and response modes, based on the nonlinear energy transfer map, provide key physical insights into practical complex self-sustained flows.
We emphasize that the interpretability of the extracted FI response and forcing modes through the energy transfer map $\mathcal{T}$ is one of the strengths of the present FI resolvent analysis.

\section{Conclusions}\label{sec:conclusion}
In this study, we proposed a forcing-informed (FI) resolvent analysis framework founded for statistically stationary unsteady flows, to aid in elucidating the underlying self-sustaining mechanisms.
Our central focus is placed on the nonlinear forcing terms in the fluctuating component of the governing Navier--Stokes equations, which are crucial for sustaining flow oscillations.
We derived the FI resolvent operator by informing the resolvent operator of the spatiotemporal property of the nonlinear forcing, specifically, the basis vectors of the input subspace spanned by the snapshots of the nonlinear forcing term, and simultaneously, the snapshots of the basis of the output subspace.
We demonstrated the capability of the FI resolvent analysis through three cases: the Stuart-Landau oscillator model, a two-dimensional cylinder wake, and a three-dimensional transitional boundary layer.
It was demonstrated that the FI resolvent operator enables the extraction of the input-output relations present in the actual flow field as the FI response and forcing modes, while the amplitudes of the output are recovered by the gain (singular values) of the FI resolvent operator.
Furthermore, based on the time-averaged energy balance between linear/nonlinear amplification/attenuation mechanisms in statistically stationary flows, we introduced a nonlinear energy transfer map that identifies the spatial domain in which the FI forcing mode supplies or attenuates fluctuation energy. 
This map aids the physical interpretations of the FI forcing mode in self-sustained oscillations.

In this paper, two numerical algorithms for computing the FI resolvent operator were proposed: a fully data-driven approach and a semi-data-driven approach. 
In the fully data-driven approach, the linear operator is estimated using the forcing snapshots, and the resulting matrix size of the FI resolvent is significantly reduced thanks to dimensionality reduction techniques widely used in the modal analysis methods.
The fully data-driven approach makes resolvent analysis of statistically stationary flows more accessible by bypassing the construction of the linear operator in a conventional way, which sometimes suffers from numerical instabilities.
Even with the proposed semi-data-driven approach for the already known linear operator, the computational cost is reduced to the same level as the randomized resolvent analysis. 
This provides an additional option to the conventional resolvent analysis for the self-sustained unsteady flows.

The basis vectors for the input and output subspaces used in the proposed FI resolvent analysis can be obtained using existing modal analysis techniques, such as spectral proper orthogonal decomposition (SPOD) and dynamic mode decomposition (DMD).
The FI response and forcing modes are expressed as linear combinations of the basis vectors, thereby ensuring that the obtained FI response and forcing modes exist in the actual flow fields.
At the same time, the temporal characteristics of the FI response and forcing modes inherit those of the basis vectors; for example, the use of the SPOD modes yields harmonic input-output modes.
Although this study employed spectral proper orthogonal (SPOD) modes for the input and output bases because of their orthogonality in Fourier space, a different choice of basis vectors with the time characteristics, e.g., a wavelet basis to analyze time-local phenomena, may also be considered.

The present FI resolvent provides key insights into self-sustaining mechanisms, whereas the conventional resolvent reveals a nontrivial linear amplification mechanism useful for flow control.
These two methods are complementary and together enhance the capabilities of the resolvent analysis framework.

\section*{Acknowledgments}
Y.I. and S.K. acknowledge the support from the JSPS Grant-in-Aid for JSPS Fellows (JP23KJ0167) and JSPS KAKENHI Grant Number 26H02169.
K.T. thanks the Vannevar Bush Faculty Fellowship (Grant Number N00014-22-1-2798).
Computer resources of the Fugaku supercomputer were provided by the RIKEN Advanced Institute for Computational Science through the HPCI System Research project (project IDs: hp240083, hp250090, hp260108).

\appendix

\section{Nonlinear forcing terms in the compressible Navier--Stokes equations}\label{app:forcingterm}
The final form of the nonlinear forcing vector in this study is given by \eqref{eq:fnl_implemtented}.
In the derivations of the nonlinear forcing terms, the $f'_{\rm{NL}}$ should be expressed as nonlinear terms with respect to the state vector $q'$; for instance, pressure is nonlinear with respect to the state vector of conservative variables in the compressible Navier--Stokes equations.
Recalling that the nonlinear term is given by $f'_{\rm NL}=F_{\mathrm{NL}}(q'^n)-\overline{F_{\mathrm{NL}}(q'^n)}$ in \eqref{eq:NLforcing}, we here derive only $F_{\mathrm{NL}}(q'^n)$ to obtain $f'_{\rm NL}$.

The compressible Navier--Stokes equations in the conservative form are written as
\begin{align}
    \frac{\partial\rho}{\partial t} &= -\frac{\partial m_j}{\partial x_j}, \label{eq:NSmass}\\
    \frac{\partial m_i}{\partial t} &= -\frac{\partial m_iu_j}{\partial x_j} - \frac{\partial p}{\partial x_i} +\frac{\partial \tau_{ij}}{\partial x_j},  \label{eq:NSmom}\\
    \frac{\partial E}{\partial t} &= -\frac{\partial (E+p)u_j}{\partial x_j} +\frac{ \partial \tau_{ij}u_i}{\partial x_j}- \frac{\partial q_j}{\partial x_j}  \label{eq:NSenergy},
\end{align}
where $m_i$ is the momentum vector $u_i$ is the velocity vector, $p$ is the pressure, $\tau_{ij}$ is the viscous stress tensor, and $q_j$ is the heat flux vector.
Here, the viscous stress tensor is given as $\tau_{ij}=\mu(T)S_{ij}$, where $\mu$ is the dynamic viscosity, determined by Sutherland's law, and $S_{ij} = \partial u_j/\partial x_i+\partial u_i/\partial x_j-(2/3)\delta_{ij}\partial u_k/\partial x_k$ under the Stokes hypothesis.
Sutherland's law is given as $\mu(T)=(T^/T_\infty)^{3/2}(T_\infty+c_1)/(T+c_1)$, where $T_\infty$ is the free stream temperature and $c_1$ is a constant (specified as $c_1=110.4/256.6$ in this study).
Thus, the viscosity is a nonlinear function of $T$, and in turn, a nonlinear function of the state vector $q$.
The heat flux vector is expressed as $q_j=-\kappa(\partial T/\partial x_j)$ where $\kappa$ is the heat conductivity.
The system is closed by the ideal gas law $p=\rho R T$, where $R$ is the gas constant.
Note that the mass conservation \eqref{eq:NSmass} does not yield any nonlinear forcing terms. 
Hence, we focus on the momentum conservation \eqref{eq:NSmom} and energy conservation \eqref{eq:NSenergy}.

To obtain the nonlinear forcing terms, we assume for simplicity that the density-fluctuation-related nonlinear forcing terms are negligibly small, that is, 
\begin{align}
    \rho' g' \approx 0.
    \label{eq:assumption_rho_fluc}
\end{align}
where $g'$ expresses the applicable fluctuating variables that appear in the governing equations, for example, $\rho', m_i' $, and $E'$.
In addition, we further assume that the fluctuations of the dynamic viscosity $\mu'$ and heat conductivity $\kappa'$, which are nonlinear functions of $q'$, are negligible.
The assumptions appear reasonable for many compressible flows, including the flows considered in this paper.  
Nevertheless, these assumptions should be carefully assessed in flows with large fluctuations in density and thermophysical properties, including high-speed jets, shock-containing flows, supercritical flows, and extremely heated or cooled wall-bounded flows.

Under the assumptions, we may obtain the nonlinear forcing terms, written as
\begin{align}
    F_{\mathrm{NL}}({q'}^n) &\approx
    \begin{bmatrix}
     0 \\
    -\frac{1}{\overline{\rho}}\frac{\partial m_i'm_j'}{\partial x_j} -\frac{\partial p'_{\rm NL}}{\partial x_i}
    \\
    -\frac{1}{\overline{\rho}}\frac{\partial}{\partial x_j}(\rho h)'m_j'
    -\frac{1}{\overline{\rho}}\frac{\partial}{\partial x_j}(\rho k)'m_j'
    +\frac{\partial}{\partial x_j}(\frac{1}{\overline{\rho}}\tau_{ij}'m_i')
    +\frac{\partial}{\partial x_j}\left( \overline{\kappa}\frac{\partial T'_{\rm{NL}}}{\partial x_j}\right)
    \end{bmatrix}.
    \label{eq:fnl_implemtented}
\end{align}
In \eqref{eq:fnl_implemtented}, the fluctuation of enthalpy $(\rho h)'$ and kinetic energy $(\rho k)'$ per unit volume are respectively given as,
\begin{align}
    (\rho h)' &= \frac{\gamma}{\gamma-1}p' + L_1(\rho'),\label{eq:enthalphy}\\
    (\rho k)' &=\frac{1}{2\overline{\rho}}(2\overline{m}_k m_k' + m_k'm_k')+L_2(\rho'),
\end{align}
where $L_i(\rho')$ expresses the linear part with respect to $\rho'$, but they will be neglected after the multiplication with other fluctuating quantities following \eqref{eq:assumption_rho_fluc}.
In \eqref{eq:enthalphy}, $p'$ is rewritten as function of $q'$ as
\begin{align}
   p'= \underset{p'_L}{\underbrace{(\gamma-1)\left(E'-\frac{1}{\overline{\rho}}\overline{m}_km_k'\right)}} 
    \quad \underset{p'_{\rm{NL}}}{\underbrace{- (\gamma-1){\left(\frac{1}{2\overline{\rho}}m_k'm_k'\right)}}}+L_{3}(\rho').
\end{align}
As shown above, the pressure fluctuation can be decomposed into the linear part $p'_{\rm{L}}$ and the nonlinear part $p'_{\rm{NL}}$ with respect to $q'$.
Using $p'_{\rm{NL}}$, the nonlinear component of temperature fluctuation is given as 
\begin{align}
T'_{\rm{NL}}=p'_{\rm{NL}} /\overline{\rho}R .
\end{align}
Finally, the fluctuation of viscous stress tensor $\tau_{ij}'$ is given as
\begin{align}
\tau_{ij}' = \overline{\mu}S_{ij}'=\frac{\overline{\mu}}{\overline{\rho}}\left(
\frac{\partial m'_j}{\partial x_i}+\frac{\partial m'_i}{\partial x_j}-\frac{2}{3}\delta_{ij}\frac{\partial m'_k}{\partial x_k}
\right).
\label{eq:viscous_stress_tensor_fluc}
\end{align}
Using these quantities from \eqref{eq:enthalphy} to \eqref{eq:viscous_stress_tensor_fluc}, the components in the nonlinear forcing \eqref{eq:fnl_implemtented} can be computed.

\section{Non-uniqueness of the basis vectors of input and output subspaces}\label{app:rankdecom}

To show the non-uniqueness of the choice of the basis vectors, let us consider the rank factorization of the forcing matrix $\widehat{F}$ defined in \eqref{eq:forcfreqmat}:
\refstepcounter{equation}
$$
\widehat{\boldsymbol{F}}(\omega) = \mathbf{\Phi}(\omega) \boldsymbol{D}(\omega)
\quad \mathrm{where}\quad  
\mathbf{\Phi}\in \mathbb{C}^{N_q\times N_\Phi}, \quad
\boldsymbol{D}\in \mathbb{C}^{N_\Phi \times N_p}.
\eqno{(\theequation)}
\label{eq:rankfact}
$$
Here, $N_\Phi$ is the rank of the matrices $\widehat{\boldsymbol{F}}$,  which is smaller than $ N_p$ (the number of the realizations) and $N_q$ (the length of the state vector). 
This is equivalent to the number of the basis vectors of the input subspaces, respectively.
From \eqref{eq:rankfact}, we clearly find that $\operatorname{Col}(\widehat{\boldsymbol{F}})=\operatorname{Col}(\mathbf{\Phi})$,  indicating that column vectors of $\mathbf{\Phi}$ are the basis vectors of input subspaces.
The non-uniqueness of the basis is proven by considering any non-singular transformation $\boldsymbol{T}$ to define a different basis as $\mathbf{\Phi}' = \mathbf{\Phi}\boldsymbol{T}$, while preserving the representation $\widehat{\boldsymbol{F}} = \mathbf{\Phi}\boldsymbol{T}\boldsymbol{T}^{-1}\boldsymbol{D}$.
The same argument applies to the basis vectors of output subspaces $\mathbf{\Psi}$ obtained via rank factorization of $\widehat{\boldsymbol{Q}}$.

Due to the non-uniqueness, the basis vectors can be either orthogonal or non-orthogonal.  
In general, the basis vectors satisfy the following weighted bi-orthonormal condition:
\begin{align}
    \langle {\phi}^{i}, {\phi}_{j}\rangle_W = \langle \psi^{i}, {\psi}_{j}\rangle_W = \delta_{ij},
\end{align}
where ${\phi}_{j},{\psi}_{j}$ are the basis vectors and ${\phi}^{j},{\psi}^{j}$ are the dual basis vectors.
For orthonormal basis vectors, the dual basis vectors are considered the same as the basis vectors, i.e., ${\phi}^{j}=\phi_j$ and ${\psi}^{j}=\psi_{j}$, which yields \eqref{eq:orthonormal}.
By introducing dual vectors, we may obtain the expansion coefficients (amplitudes) for these non-orthogonal basis vectors, which are used in $\boldsymbol{T}_\Phi$.
This discussion extends the choice of modal decomposition methods for basis estimation, such as DMD, in which the modes are not always orthogonal, as discussed in \S\ref{sec:modalanalysisforbasis}.

\section{Nonlinear energy transfer maps from FI response and forcing modes}\label{app:energymap}
\begin{figure}
    \centering
    \includegraphics[width=0.9\linewidth]{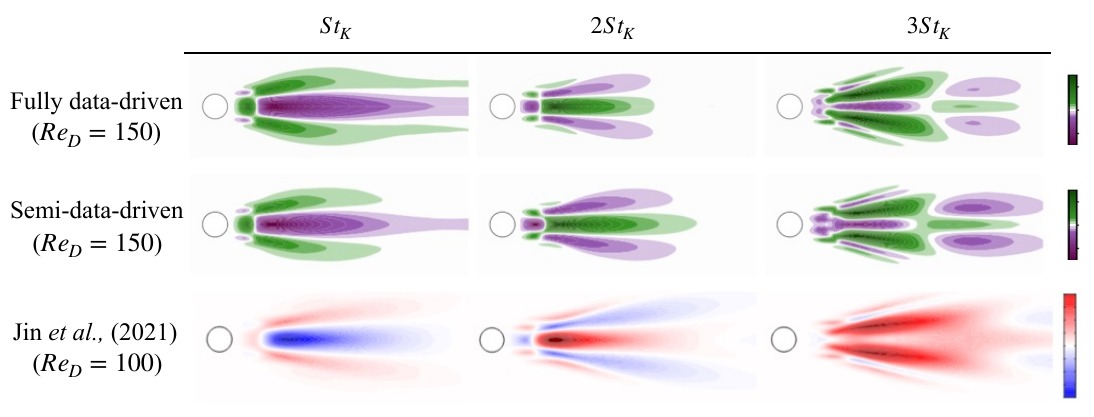}
    \caption{
    Comparison between the present nonlinear energy transfer maps computed from the most dominant FI response and forcing modes  (top: fully data-driven approach; middle: semi-data-driven approach) and the map based on DNS data (bottom) in \citet{Jin2021-ge} (Adapted with permission from \citet{Jin2021-ge}. Copyright (2021) by American Physical Society).
    }
    \label{fig:vsJin}
\end{figure}

We show that the nonlinear energy transfer maps correspond to the one proposed in \citet{Jin2021-ge} in Fig.~\ref{fig:vsJin}.
In this figure, the nonlinear energy transfer map for summation of the all variables ($\rho',m_x',m_y',E'$), that is, $\mathcal{T}_{all}=\mathcal{T}_{m_x}+\mathcal{T}_{m_y}+\mathcal{T}_{\rho}+\mathcal{T}_{E}$ is shown, where we define $\mathcal{T}_{m_y},\mathcal{T}_{\rho},$ and $\mathcal{T}_{E}$ in a same manner as $\mathcal{T}_{m_x}$ in Fig.~\ref{fig:mode_data_cylinder2D}(bottom).
We have confirmed that $\mathcal{T}_{\rho}=0$ since no nonlinear forcing appears in the mass conservation as in \eqref{eq:fnl_implemtented}.
We observe qualitative agreement among the nonlinear energy transfer maps from the preset fully data-driven approach, the map from the semi-data-driven approach, and the map computed using DNS data in \citet{Jin2021-ge}.
The differences may be attributed to the Reynolds number and the governing equations (we employ the compressible NSs, whereas \citet{Jin2021-ge} consider the incompressible NSs).
\cite{Jin2021-ge} discussed that the conventional resolvent analysis fails to predict the nonlinear transfer map and pointed out the requirement of the information of the inter-scale nonlinear interactions for a better prediction using harmonic resolvent analysis \citep{Padovan2020-jx}.
As the nonlinear interactions are considered, the present FI resolvent analysis successfully captures the nonlinear energy transfer map.

\bibliographystyle{unsrtnat}
\bibliography{ref}

\end{document}